\def \myplotone#1 {\plotone{#1}}
\def \myplotfiddle#1 {\plotfiddle{#1}{512pt}{0}{100}{100}{-288}{-140}}
\def\myplottwo#1#2{\centering \leavevmode
\epsfxsize=.42\columnwidth \epsfbox{#1} \hfil
\epsfxsize=.42\columnwidth \epsfbox{#2}}

\def \NgalI {736}
\def \NgalV {690}
\def \NobjI {2148}
\def \NobjV {1757}
\def \Imagcut {20.5}
\def \Vmagcut {22}
\def \Vcomplim {24.4}
\def \Icomplim {23.0}
\def \et {{\it et al. }}
\def \etal {{\it et al. }}
\def \Mpc {h^{-1}{\rm Mpc}}
\def \kpc {h^{-1}{\rm kpc}}
\def \farcs{\hbox{$.\!\!^{\prime\prime}$}}
\def \farcm{\hbox{$.\!\!^{\prime}$}}

\documentstyle[aaspp]{article}

\eqsecnum

\slugcomment{CfPA 96-th-09/Submitted to Ap.J.}

\begin{document}
\title{A Weak Gravitational Lensing and \\X-ray Analysis of Abell~2163}
\author{G. Squires\altaffilmark{1},
D. M. Neumann\altaffilmark{2},
N. Kaiser\altaffilmark{3}\altaffilmark{,7},  
M. Arnaud\altaffilmark{6},
A. Babul\altaffilmark{4}, 
H. B\"ohringer\altaffilmark{2},
G. Fahlman\altaffilmark{5}\altaffilmark{,7}, 
and D. Woods\altaffilmark{5}
}
\altaffiltext{1}{Center for Particle Astrophysics, University of California, Berkeley, CA, USA 94720}
\altaffiltext{2}{Max-Planck-Institut f\"ur extraterrestrische Physik,
D-85740 Garching bei M\"unchen, Federal Republic of Germany}
\altaffiltext{3}{Canadian Institute for Advanced Research and
Canadian Institute for Theoretical Astrophysics, University of Toronto,
60 St.\ George St., Toronto, Ontario, Canada M5S 1A7}
\altaffiltext{4}{Dept. of Physics, New York University,
4 Washington Place, Rm 525, New York, NY, USA 1003-6621}
\altaffiltext{5}{Dept. of Geophysics and Astronomy, University of
British Columbia, 2219 Main Mall, Vancouver, BC, Canada V6T 1Z4}
\altaffiltext{6}{Service d'Astrophysique, CEA Saclay, 91191 Gif-sur-Yvette, 
France}
\altaffiltext{7}{Visiting Astronomer, Canada-France Hawai'i Telescope.
Operated by the: National Research Council of Canada, le Centre National
de la Recherche Scientifique de France and the University of Hawai'i}

\begin{abstract}
We report on the detection of dark matter in the cluster of galaxies
Abell~2163 using the weak gravitational distortion of background galaxies, 
and an analysis of the X-ray emission
{}from the cluster. We find that while the qualitative distributions of the
cluster light and the dark matter are similar --- shallow and extended, with 
significant substructure --- the X-ray morphology shows a more regular overall
appearance. We interpret the joint lensing and X-ray 
observations as a signature of a merger event in the cluster. 
We present new ROSAT/HRI data and reanalyze ROSAT/PSPC data,
accounting for the effect of a varying background to determine the best fit 
parameters in the $\beta$-model formalism. We combine the surface brightness 
fits with two determinations of the radial temperature profile to determine 
the total mass. Although there are slight variations in the total mass 
determinations introduced by the uncertainties in the $\beta$-fit, the main 
contributor to the error arises {}from the uncertainties in the temperature 
determinations.  Even though the morphologies of the dark matter/light and 
X-ray gas are quite different, we find that the total mass determined 
{}from the X-ray and weak lensing estimates are consistent with each other 
within the $2\sigma$ error bars, with the X-ray inferred mass a factor of 
$\simeq 2$ larger. However, as the lensing mass estimates are differential 
(the surface density at any point is determined relative to the mean in a 
control annulus), the shallow, extended nature of the mass profile biases the 
lensing inferred mass downwards. We estimate the correction for this effect 
and find very good agreement between the 
corrected lensing and X-ray results. We determine the gas mass fraction in 
this cluster and find $f_g \simeq 0.07 \;h^{-3/2}$ at all radii and a constant 
mass-to-light ratio of $M/L_V = (300 \pm 100) \;h M/L_{\odot V}$. 
\end{abstract}


\section{Introduction}

Determining the masses of the visible and the dark components of the 
galaxy clusters and mapping out their relative distributions is the
key towards understanding  both the dynamical state of the clusters and      
their evolutionary history. As the largest clearly defined objects in the
Universe, galaxy clusters are important cosmological probes. 
They are used, for example,
to measure the matter composition of the Universe - e.g., the luminous
baryon to dark matter ratio (\cite{white93}) or the mass-to-light ratio
(e.g., \cite{blumenthal84}) on large scales. They are also
important tracers of the large scale structure as measured
through the mass function or the spatial correlation function (e.g.,
\cite{efstathiou95}). 

Ideally, one would like to study X-ray bright 
clusters whose galaxies population has been spectroscopically studied 
and that also induce measurable, even if only weak, gravitational
distortions in the images of faint background galaxies.
The X-ray emission of the hot intracluster 
plasma provides an attractive method to determine cluster masses
(\cite{sarazin86}; \cite{hughes89}). The derivation
of the gravitational mass using X-ray data rests on the assumption,
however, that the gas is in hydrostatic equilibrium in the cluster
potential. Since the dynamical time scale for the 
formation of clusters is comparable to the Hubble time and, as many 
clusters are found to have substructure implying that the are dynamically 
young and unrelaxed,  the X-ray mass derivations are not without
uncertainties.

The discovery of gravitational lensing effects, both strong and
weak, in galaxy clusters opened
a new way to probe the gravitational potential of clusters,
which is free of any assumption of the cluster dynamical state  
(e.g., \cite{fort94}). Strong lensing effects resulting in obviously distorted
images of background galaxies -- observed as arcs -- were first used 
to obtain cluster masses by modeling the lens effects. However,
such analyses are highly model dependent and are limited to the central
regions of the cluster.

The analyses of the weak gravitationally induced
distortions in the images of faint, background galaxies, 
on the other hand,  offer a unique opportunity to directly
probe the total mass distribution (\cite{tyson90}; \cite{ks93}; 
see also \cite{squires96} and references therein).
The weak lensing effects can be measured and inverted to derive the mass
distribution in the cluster in a model independent way and free of any
assumption of the cluster symmetry and dynamical state. This permits both
a non-parametric determination of the total mass, and
a 2D map of the total mass distribution in the cluster.

Taken together,
the weak lensing analysis and an analysis of X-ray observations offers a
unique possibility to probe the relative distributions of the gas and the
dark matter and study the dynamical relationship between the two.
It is worth noting that, on a case-by-case basis, discrepancies in the
masses determined by the two methods can naturally arise
(for example, if the cluster is strongly elongated cluster along the line 
of sight). Thus, to be able to make general conclusions about the 
cluster population, the analyses need to 
be compared for several clusters for which both good optical and X-ray
data are available. 

Miralda-Escud\'e \& Babul (1995) (hereafter MB) performed the first joint
X-ray and lensing study 
of the clusters A2218, A1689 and A2163. In the former two clusters they found 
that the mass in the central region
implied by the observed giant arcs is a factor of 2-2.5 larger
than the mass derived {}from X-ray data if the intracluster plasma is
assumed to be isothermal at the observed temperature 
and in hydrostatic equilibrium in the cluster
potential. A similar discrepancy, but not quite as strong, was 
confirmed by Kneib \etal (1995) also for A2218 in a more detailed effort to
model of the strong lensing effect. In contrast, the
mass determined from the strong lensing in the regular cluster PKS0745-191 is 
in agreement with the mass derived from X-ray data, using a 
multiphase cooling flow model.
If it is a common occurrence, the X-ray/lensing mass discrepancy could have 
important implications 
for quantities, such as the cluster gas fraction $M_{\rm gas}/M_{\rm tot}$, 
derived solely {}from X-ray data.  Typically, the cluster gas fraction 
is estimated to be $M_{\rm gas}/M_{\rm tot} \geq 0.05 \;h^{-3/2}$ 
(\cite{white95}; \cite{david95}) a result that has been a source of 
much discussion (\cite{white92}; \cite{babul93}; \cite{white93}). It is worth
noting, however, that the lensing mass determinations
based on the strong lensing features are, however, 
restricted to probing the central regions of clusters.

More recently, Squires \etal (1996) used the weak gravitational lensing 
distortions to map the mass distribution in A2218 out to a radius of
$\sim 0.5 \Mpc$, and jointly analyzed the X-ray, optical and weak
lensing results.  Under the assumption that the cluster mass distribution
extends out to $\sim 1 \Mpc$, as indicated both by the lensing mass
profile and the X-ray surface brightness profile, the results suggested 
that the lensing/X-ray mass discrepancy of kind found by MB
may extend systematically beyond cluster core 
(although the combination of the uncertainties in the lensing and
X-ray mass estimates, the latter largely
due to poorly determined cluster temperature profile, 
did not preclude agreement between the two estimates).
This discrepancy was recently confirmed by Loewenstein (1996) for both 
A2218 and
A1689 using new ASCA determinations for the
temperature profiles. Conversely, in a similar study of the clusters
MS1455 and MS0016, Smail \etal (1995) found that 
the lensing and X-ray masses are in agreement (although we note
that the X-ray analyses in
these latter cases were based on very sparse X-ray spectroscopic
data and thus the uncertainties on the X-ray mass determinations were
larger). 

In this paper, we present a study of the galaxy, gas, and gravitational
mass distribution in the cluster of galaxies 
A2163 ($z=0.201$). This cluster is 
the hottest cluster and one of the two most massive galaxy clusters
known so far (\cite{arnaud92}). 
Based on GINGA satellite measurements, A2163 has an X-ray
temperature of $\sim$ 14~keV and a X-ray luminosity of 
$6 \times 10^{45}$~erg~s$^{-1}$ (\cite{arnaud92}).
It is one of the rare clusters for which the X-ray emission has been         
traced out to a distance similar to the virial radius; the 
emission was detected         
significantly with the ROSAT/PSPC up to $2.3 \Mpc$ or 15 core radii 
(\cite{elbaz95} -- hereafter EAB). 
While the total mass within that radius, derived {}from
the GINGA and PSPC data, is exceptionally high (2.6 times greater than
the total mass of Coma), the corresponding gas mass fraction, 
$\sim 0.1 \; h^{-3/2}$, is typical of other rich clusters.  A quick drop of the
temperature at large radii ($\sim$ 4~keV beyond $6$ core radii) was
observed recently with ASCA, strongly constraining the total mass
profile, assumed to follow a simple parametric law (\cite{MMIYFT95} --- 
hereafter referred to as MMIYFT). 
The temperature distribution in A2163 also shows evidence for complex 
non-azimuthally symmetric temperature variations in the central 
regions (Markevitch \etal 1994).
A2163 exhibits the
Sunyaev-Zel'dovich effect (\cite{wilbanks94}) and is remarkable in the
radio, having the most luminous and extended  halo yet detected
(\cite{herbig95}).  The spatially resolved  measurements of the the
Sunyaev-Zel'dovich effect, combined with  the PSPC data, also confirm the
decrease in the temperature in the outer part of the cluster 
(\cite{holzapfel96}).

In spite of its exceptional X-ray and radio properties, the 
{\em dynamical} state of A2163 is puzzling. The high 
mean X-ray temperature and high X-ray luminosity would suggest a 
massive cluster with a very deep potential well that ought to contain 
a plethora of strong lensing features such as arcs and arclets.  In 
fact, only two arcs have been observed and they lie at a relatively 
small distance {}from a brightest cluster galaxy. This is very much
in contrast to the lensing features detected in other hot clusters 
(e.g., A2218).  The optical properties of A2163 are also quite unassuming
in comparison to compact clusters such as A2218 and A1689.  It is 
classified as an Abell richness class 2 cluster, its central
galaxy is not a cD galaxy, and the cluster galaxy distribution is  
irregular and extended.  On the other hand, A2163 has a very
high velocity dispersion ($\sigma=1680$~km/s) and a flat, very
extended velocity histogram (Soucail \etal 1996; Arnaud \etal~1994).
The optical data,
together with the detailed X-ray morphology, have been interpreted as signature
of a recent or ongoing merger of two large clusters (EAB; Soucail \etal 1996).

For the present study, we acquired optical images in V- and I-bands
using the Canada-France-Hawaii telescope. The total mass distribution of
A2163 is determined {}from the distorted images of the faint background
galaxies  using the algorithm of  Kaiser \& Squires (1993) and
amendments (\cite{sk95}). We also derive the cluster galaxy light and
surface number density distribution.
We reanalyze the ROSAT/PSPC data of EAB, allowing for                 
statistical uncertainties in the background, and utilize
ASCA temperature information provided by MMIYFT.
Furthermore we present new ROSAT/HRI data, which
resolves the center of the cluster to a much higher accuracy
than the PSPC data. We reinvestigate the total mass and gas mass
fraction  determination based on X-ray data
using the method developed by Neumann and B\"ohringer
(1995) which is free of any assumption on the gravitational potential
shape, in contrast with parametric methods used in previous studies of
this cluster. We make qualitative comparisons of the 
morphology of the gas, galaxy and mass distribution 
and give new insight into the dynamical state of
the cluster. We compare the X-ray and
lensing total mass estimates, determine the baryon fraction as well as the
mass-to-light ratio profile. Finally we consider the cosmological
implications of our results.

\section{Optical Data Acquisition and Analysis}

A2163 was observed using the 3.6m Canada-France-Hawai'i telescope
on the nights of 1994 June 6-9.  The detector used was the 2048~x~2048
Loral 3 CCD at prime focus with a pixel size of 0\farcs207.  
The observations of A2163 comprised of $3 \times 20$~minute exposures 
in V and I, with seeing of 0\farcs80 and 0\farcs85 (FWHM) in V and I
respectively.
We observed the central 7$^\prime$ of the cluster
covering a square of side $\simeq 1\Mpc$ at the cluster redshift. 

A detailed description of the  data reduction procedure for this data sample 
is given in 
Squires \etal (1996). Briefly, each bias subtracted image was divided 
by a median twilight flat. The residual rms scatter in the sky background on
the individual images was 24.1 magnitudes per 
square arcsecond in I, and 25.2 magnitudes per square arcsecond in V.
The data was calibrated against photometric standards in the globular
clusters M92 and NGC 4147 (unpublished photometry {}from \cite{davis90};
see also \cite{stetson88} and \cite{odewahn92}) and 
Landolt (1992) standards in SA110. Color terms were found to be
unnecessary in the transformation and the I and V zero points were
determined with a formal error of less than 0.005 mag.
Using the summed V- and I-band images, 
we detected $\NobjV$ and $\NobjI$ objects in V and  I respectively,
with a significance of $\nu \geq 4\sigma$ over the local sky 
background. The catalogues are complete to a limiting magnitude of
$\Vcomplim$ in V and $\Icomplim$ in I.
We measured the position, shape and brightness parameters using our 
standard procedure (\cite{kaiser95}) as described in Squires \etal (1996).

\subsection{Cluster Luminosity and Light Distribution}

To obtain a qualitative description of the cluster light distribution,
we calculated the cluster galaxy light and surface number density. 
We combined the V and I observations to determine the galaxies' color
and identified a sequence of red objects
with $V-I \simeq 1.5$ (color corrected for reddening by
E(V-I) = 0.2; see \cite{bessell88}), which we classified as
candidate cluster galaxies. 
To extract this color-selected subsample, we fitted a linear model to the 
color sequence and identified bright objects $(I<21)$ with color within 
0.2 magnitudes of the mean.

\begin{figure}
\myplottwo{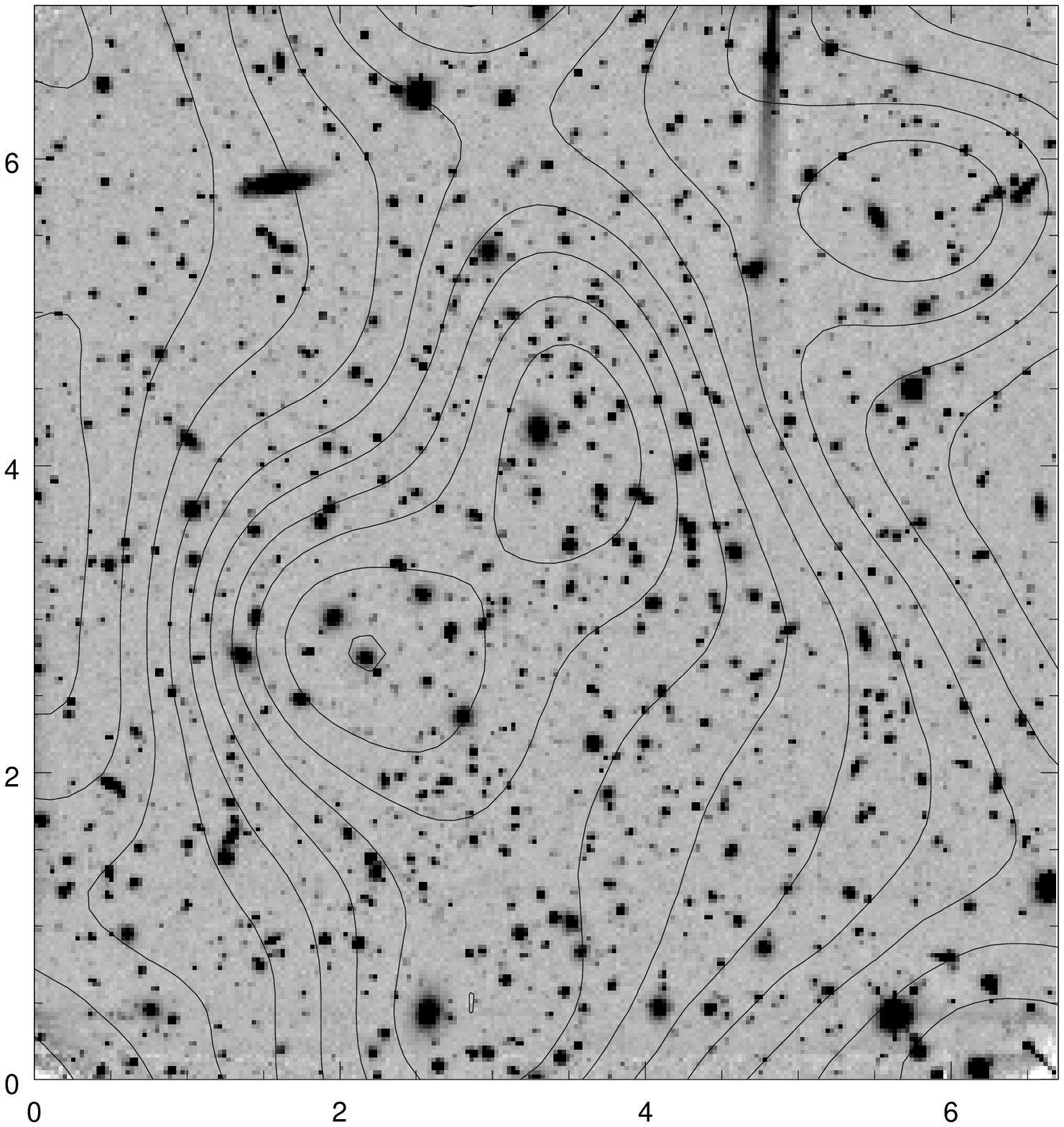}{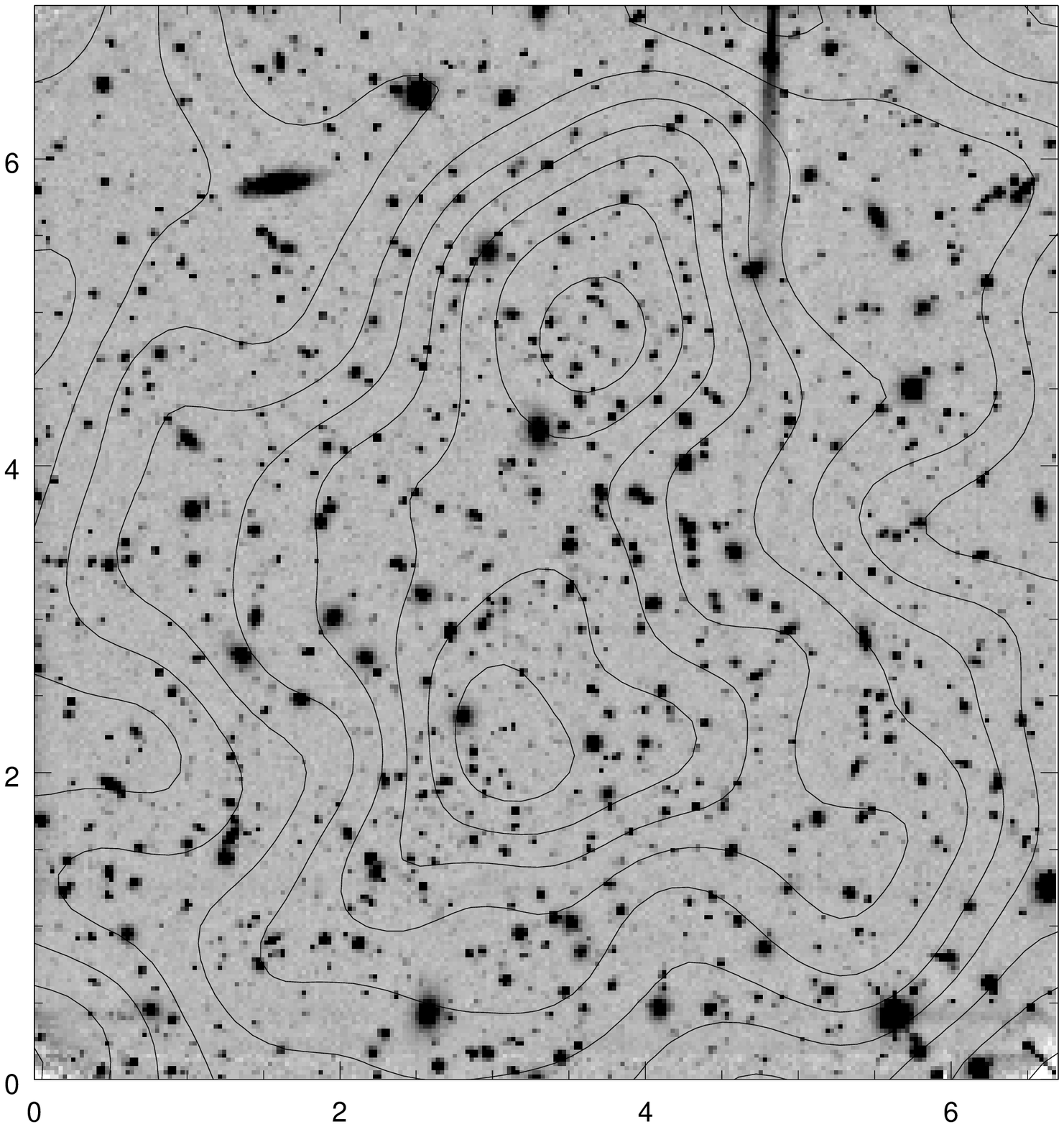}
\caption{The luminosity and number weighted contours of the color-selected 
cluster galaxy sample placed on the rebinned
V-band image of the cluster. 
The left plot shows the luminosity distribution and the right panel is 
the galaxy number distribution. The contours have been smoothed
with a Gaussian filter with scale 0\farcm66. 
North is to the right: East is up. The angular scale is arcminutes
$(1^{\prime} = 0.127 \Mpc)$.}
\label{fig:a2163_lightdistributions}
\end{figure}

We display 
the color-selected cluster galaxy light distribution and surface 
number density as contour plots placed on the rebinned V-band image of the 
cluster in Figure \ref{fig:a2163_lightdistributions}. The
contours have been smoothed with a Gaussian filter with scale 0\farcm66. 
The galaxy light distribution shows three peaks: one centered
on a giant elliptical galaxy near the center of the frame ---
this is  the galaxy that has two associated arcs 15\farcs6 to the 
west. Two other
peaks in the galaxy light distribution appear to the northeast and southwest
of this galaxy, and there is an extension towards the western edge of the 
image.  The galaxy number density distribution is broadly similar, with
the main peak slightly displaced {}from the central giant elliptical
galaxy and a second peak about 2\farcm5 to the west. Together, these
plots give the impression of a somewhat irregular and extended cluster --- 
the relative
strengths of each light concentrations are comparable. 
Qualitatively, this is also apparent {}from a
visual inspection of the image: there are several prominent 
luminous galaxies
distributed over the frame, with no obvious dominant galaxy defining an
optical center.
Spectroscopy of this field confirms the very flat and extended nature of the
galaxy distribution in this cluster (\cite{arnaud94}; 
\cite{soucail96}).

\begin{figure}
\myplotone{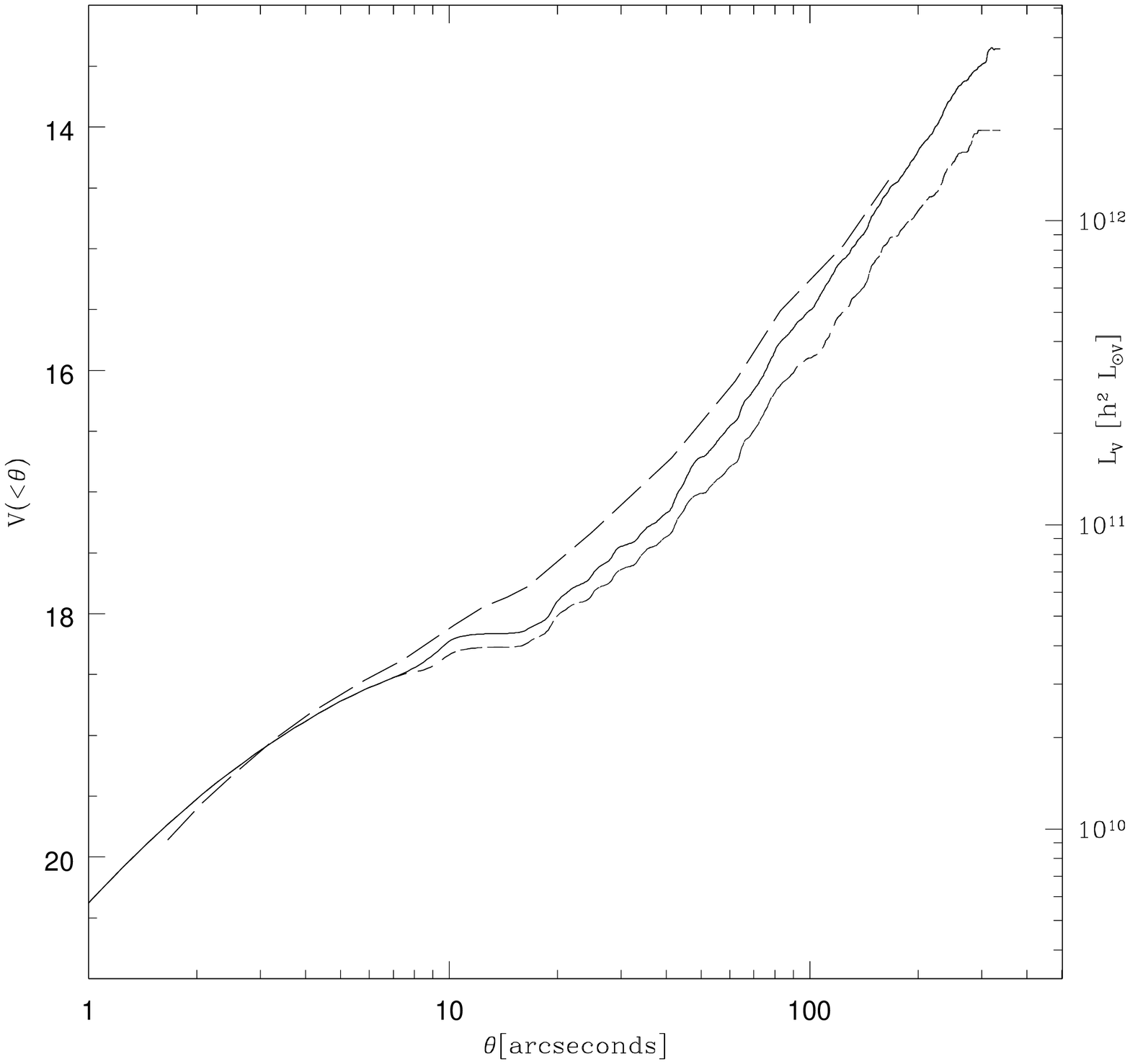}
\caption{The cumulated V-magnitude and $L_V$ luminosity for A2163.
The solid lines
come {}from adding the light {}from all galaxy candidate objects in the
field. The short-dashed line is for the color-selected sequence and the
long-dashed line is the `total' light in the cluster calculated by adding
all the light on the image after masking cosmic rays and stars.
The scale is $60^{\prime\prime} = 0.127 \Mpc$.}
\label{fig:LightProfiles}
\end{figure}

We formed three estimates of the cumulative light profile in the cluster,
taking the giant elliptical galaxy as the cluster center.  The light profiles
are displayed in Figure \ref{fig:LightProfiles}.
The short-dashed line shows the result of considering only the light 
associated with the color-selected galaxy sequence. The solid-line shows 
the profile obtained by adding the light, as determined {}from aperture 
magnitudes, associated with {\em all} the galaxies detected in the image. 
To account for
the possibility that there is extended intracluster light present, or light in
extended galaxy halos, we also
calculated the total light directly {}from the images, masking 
only cosmic rays and
stars and adding all the light {}from the cleaned image. 
The apparent magnitudes  associated with the galaxies
were converted into luminosity using the transformation 
$L_V = 10^{ 0.4 ( M_{V \odot} - V + DM + K + A_v) }L_{\odot}$,
where $M_{V \odot} = 4.83$ is the total solar V magnitude and 
$DM = 39.00$ is the distance modulus for A2163 in $\Omega = 1$, $h = 1$ 
cosmology (in computing the light profiles, we assumed that all sources were
at the cluster redshift).
We also applied a K-correction of $K = 0.4$ as suggested by the 
data in Coleman, Wu and Weedman (1980) and used an extinction correction 
of $A_v = 0.54$. 

The resulting 
profile is shown as the long-dashed line.  In spite of the fact that the 
total light profile exceeds that associated with the cluster sequence 
subsample by $\simeq 30$\% (the total light profile also includes contribution
{}from background and foreground galaxies),  all three estimates 
exhibit similar
trend as a function of radius.  At angular radius $\theta > 1^\prime$
{}from the central elliptical galaxy, the cumulative light profiles rise as 
$\theta^{1.6}$, implying that the azimuthally-averaged
surface brightness profile of the cluster is $l(\theta)\propto \theta^{-0.4}$.
This surface brightness profile is shallower than isothermal 
($\propto \theta^{-1}$).

\subsection{Lensing Analysis} \label{sec:lensing}

Since the seeing conditions for both the I- and V-band observations were 
similar, we employed both data sets for the lensing analysis.  The I-band 
images do have somewhat higher sky noise than the V-band images 
and hence, the corresponding results are slightly noisier.

We selected as candidate
background galaxies all objects with half-light radius greater than
1.2 times the mean stellar half-light radius and applied a bright
magnitude cut  ($V< \Vmagcut$ and $I< \Imagcut$) to reduce foreground 
and cluster galaxy
contamination. This cut is somewhat arbitrary --- the values quoted
above were chosen after some experimentation to maximize the signal-to-noise 
in the reconstructions (although the change is rather small if the above
cuts are increased by $\simeq 1$ magnitude). With 
these selection parameters, we found 
$\NgalI$ galaxies in the I-band data and $\NgalV$ in V.

Before converting
the measured galaxy shapes into estimates of the cluster
surface density, we corrected for two effects: The anisotropy
in the point spread function (psf) due to wind-shake, guiding errors, etc.,
and the smoothing due to seeing. The procedure to correct for these
systematic errors is described fully in our analysis of 
A2218, (\cite{squires96}). Briefly, we
mapped the stellar psf across each image and applied an empirical correction 
(Kaiser, Squires \& Broadhurst 1995) to each object to remove the anisotropy
in the psf. We found a residual stellar ellipticity of less than 0.5\% 
after the correction had been applied. Similarly, the
effect of seeing was calibrated by simulations using the HST-MDS data.
We applied a constant shear to the images and
created a simulated data set that had the same pixel 
resolution, exposure time, 
sky noise, and seeing conditions as the data observed at CFHT. 
We then processed the 
simulated data in an identical fashion to the CFHT data to determine
the signal loss due to seeing. We repeated this simulation a number of times,
applying different shear and random sky noise realizations and
averaged the signal recovery fraction.

\begin{figure}
\myplottwo{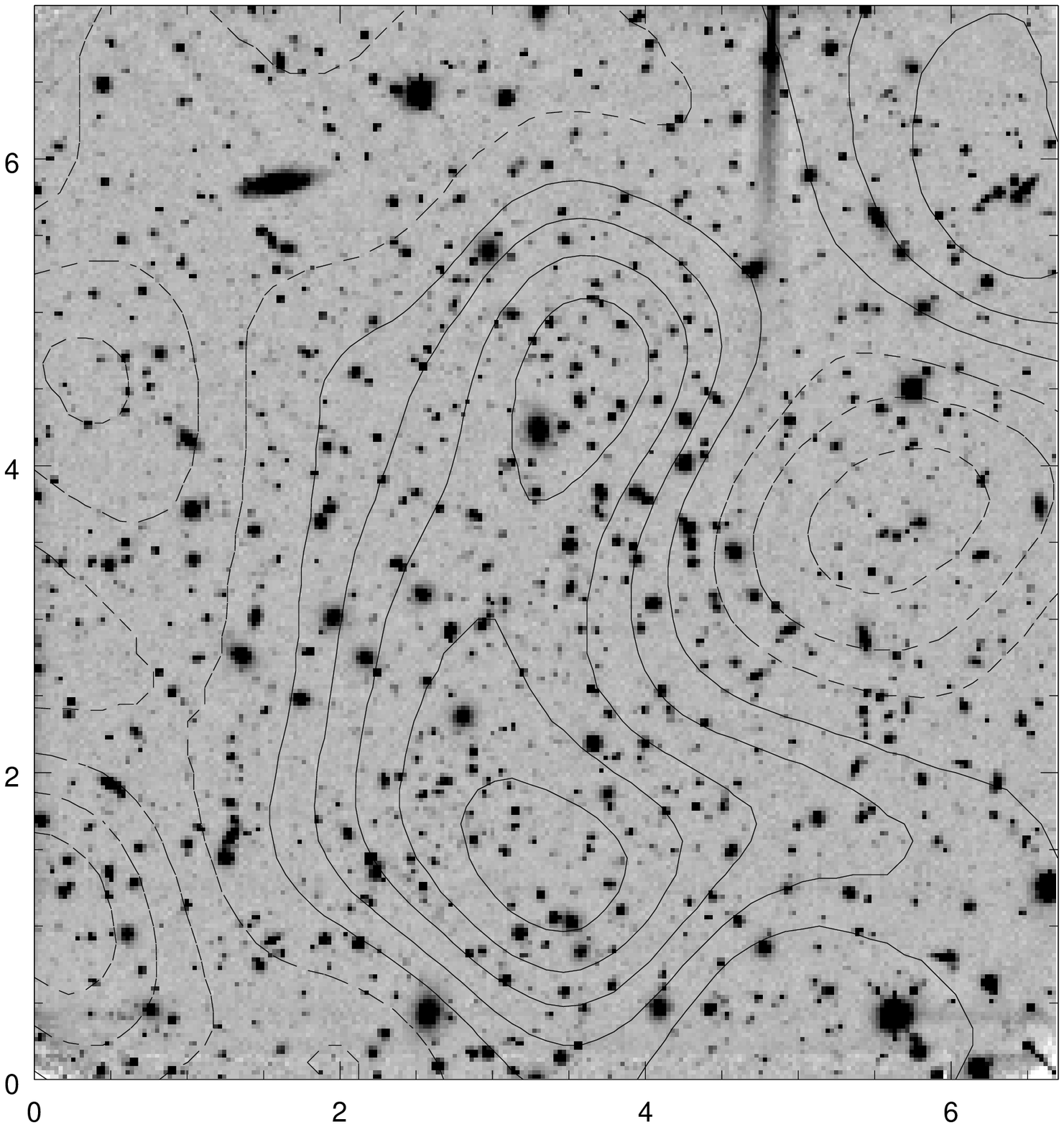}{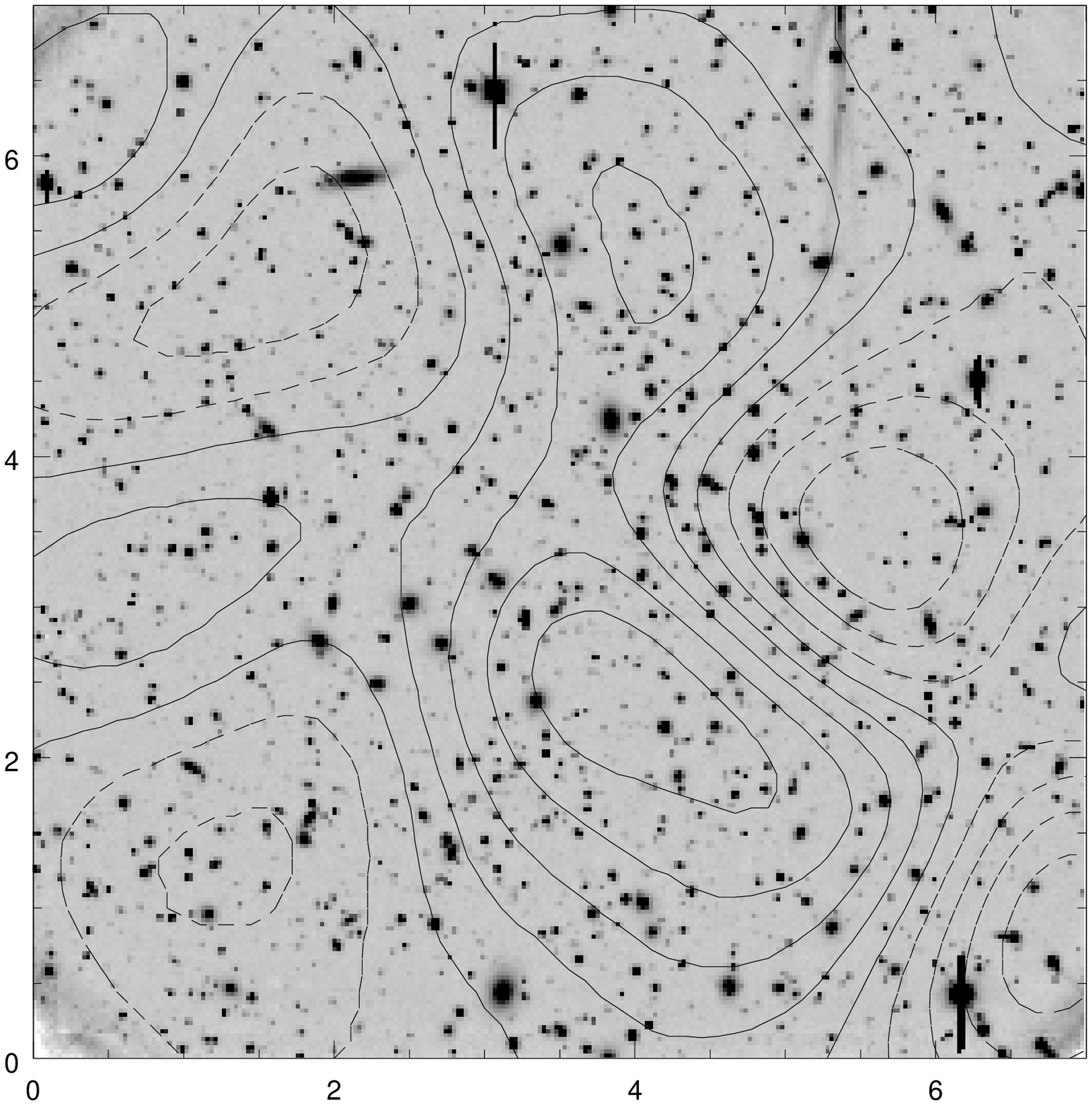}
\caption{The reconstructed surface densities placed on the rebinned V-band
image of the cluster. The V- and I-band results are shown in the left/right
panels respectively. The contours have been smoothed with a 
0\farcm66 Gaussian filter. North is to the right; East is up and the angular
scale is arcminutes $(1^{\prime} = 0.127 \Mpc)$.}
\label{fig:a2163_massonimage}
\end{figure}

The seeing and psf-corrected galaxy shape measurements were converted into
the cluster's surface mass density distribution using
the maximum probability extension (\cite{sk95}) 
to the original Kaiser \& Squires (1993) algorithm. The calculation used 15
wavemodes and a regularization
parameter of $\alpha = 0.05$ (changing this value by factors of $0.5$ to $
2\times$
changes the amplitude of the results by $\simeq 4$\%).
In Figure \ref{fig:a2163_massonimage} we display the mass surface
density reconstruction for Abell~2163.  The results for both the 
V- and I-band data are shown.

\begin{figure}
\myplotone{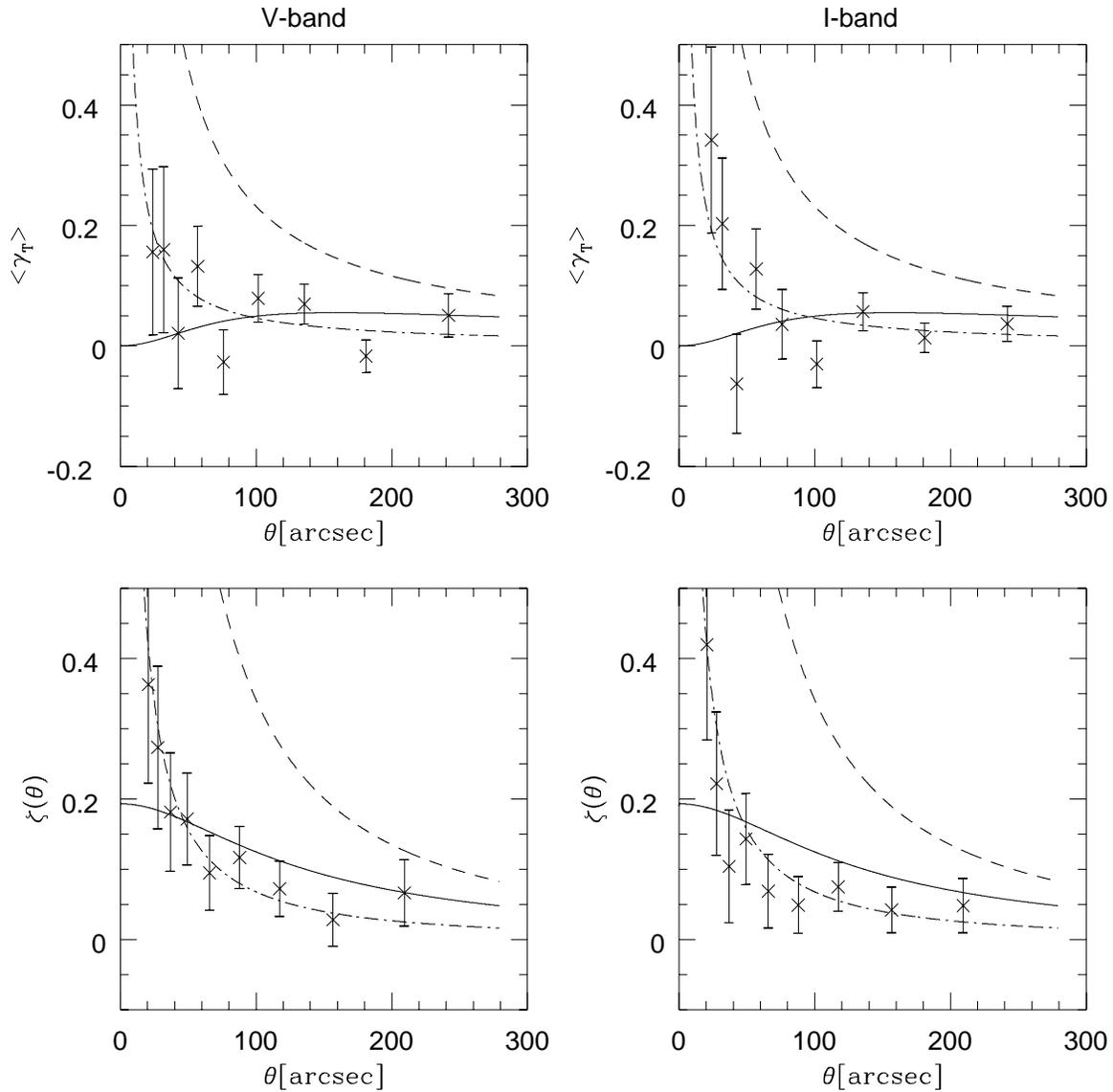}
\caption{The top panels show the azimuthally-averaged, mean radial
shear about the giant
elliptical galaxy centroid. The bottom panels show the $\zeta$ statistic. 
The V and I-band
results are shown in the left and right columns respectively.
The solid line is the prediction for a softened isothermal model 
with the measured velocity dispersion of $\sigma = 1680$~km/s and a
core of 1\farcm2 (which corresponds to the X-ray core radius).  
The dashed line
is the prediction in the singular isothermal model with the same
velocity dispersion. The dot-dashed line is the best-fit singular
isothermal model ($\sigma = 740$~km/s). The scale is 
$60^{\prime\prime} = 0.127 \Mpc$. }
\label{fig:a2163_etprof}
\end{figure}

The contour maps of the mass distribution in Abell~2163 is rather irregular 
and quite different {}from the more symmetric distributions that we have seen
in other clusters (eg. A2218 --- see Squires \etal 1996).  The V-band mass map
shows two mass peaks, with possibly a third one in the northeast corner of 
the image.  One of the mass
peaks appears to be associated with the giant elliptical galaxy that has
two arcs nearby.  The significance of each of the peaks, however, is 
rather modest
$(\leq 3\sigma)$ and is lower than that we have found for other clusters such
as A2218.  This is not surprising given the small number of galaxies used in 
the reconstructions (N$\simeq 700$) and the moderate seeing conditions.
Furthermore, the amplitude of the shear is quite small 
(see Figure \ref{fig:a2163_etprof}) which naturally
arises if the surface mass distribution is flat. 
It does, however, mean that the precise location of the peaks should be 
interpreted with caution.  The I-band mass map also features two main mass
peaks and like the V-band mass map, shows that the cluster mass distribution 
is elongated primarily along the East-West direction, with two main mass 
concentrations.  We attribute the small scale differences in the 
V- and I-band mass maps to a combination of mass peaks of modest significance
and the higher sky noise affecting the I-band data. (It is possible that the 
eastern peak with northern extension that appears in the I-band data is 
the combination of the eastern and northeastern peaks that appear in the 
V-band).  Taking the average of the two maps yields a result that is 
smoother, but
still shows the bimodal structure, again with only moderate
statistical significance (the ``averaged'' 
peaks have significance of $3\sigma$ and $2.5\sigma$ respectively).

The small scale fluctuations between the V- and I-band results noted above
have little impact on the global statistics, such as the gravitational shear
profile and the cluster radial mass profile, that we are interested in and 
which we can determine with higher significance.  We display
in Figure \ref{fig:a2163_etprof} (top panels) the azimuthally-averaged, 
mean radial shear profile for A2163 computed about the 
giant elliptical galaxy with associated arcs.  We have also done 
the calculations with the centroid of the X-ray emission chosen as the 
cluster center in order to facilitate comparisons with results derived
{}from the X-ray observations.  The results within 
$\theta \simeq 1^{\prime} \simeq 130 \kpc$ should be interpreted with 
care;  any contamination by the cluster (or foreground) galaxies 
of the catalog of faint galaxies used to determine the shear profile  
will tend to dilute the signal and lower the  measured shear. The problem 
is of particular concern in the present case due to the small number of faint
galaxies available for determining the gravitational shear.  Fortunately,
the contamination only tends to affect shear measurements for the central
regions in the cluster.
In our analyses of A2218 (Squires \etal 1996) and MS~1224 (\cite{fahlman94}), 
we found that the contamination mostly affected results for the innermost 
$\sim 100 \kpc$, but is negligible at larger radii.

For comparative purposes, we also show in Figure \ref{fig:a2163_etprof} 
the expected shear for a softened isothermal-like surface density 
distribution,
$\Sigma(\theta) \propto \sigma^2 (\theta^2 + \theta_c^2)^{-1/2}$, 
where $\sigma$ is the line of sight velocity dispersion and
$\theta_c$ is a softening (core) parameter.  We consider three cases: In two
of the cases, we use the measured
velocity dispersion $\sigma = 1680$~km/s and show the results for a singular
distribution, $\theta_c = 0$ (dashed curve), and a distribution where the
core radius is set equal to that of the cluster's X-ray surface brightness 
distribution, $\theta_c = 1\farcm2$ (solid curve).  In comparing the observed
and the model shear profiles, it is clear that the observed shear profile is 
much weaker than that induced by a singular
isothermal profile characterized by a velocity dispersion comparable to that
observed for the cluster galaxies, even if one takes into consideration the
possibility that shear in the inner $1^\prime$ might be higher than observed.
On the other hand, a mass distribution 
also characterized by $\sigma=1680$~km/s but with a core radius comparable
to that observed in the X-ray produces less shear than observed, particularly
in the inner $1^\prime$.  In fact, such a profile would not be able to 
produce the thin large arcs observed in the vicinity of the giant elliptical
galaxy.  Even if the cluster surface mass is flat over
the range $1^\prime < \theta < 2^\prime$, both the observed shear profile and
the thinness of arcs suggest that the cluster mass profile must steepen in
the central regions.

We also consider a singular model that provides a reasonable fit to the 
measured shear (dot-dashed curve). This model corresponds to $\sigma=740$~km/s
although a singular isothermal model with $\sigma$ as high as $1000$~km/s
is also consistent with the observations, particularly if one takes into 
account the possible suppression of shear in the inner regions of the cluster
by contamination.  Indeed, under the simple assumption of a spherical,
isothermal mass distribution, a velocity dispersion of $\simeq 930$~km/s
would be required to form the Einstein radius at the observed arc
position.

In the bottom panels of Figure \ref{fig:a2163_etprof}, we display the 
statistic (Kaiser \etal 1995)
\begin{eqnarray}
\zeta(\theta_1, \theta_2)& =& 2( 1 - \theta_1^2 / \theta_2^2 )^{-1}
	\int_{\theta_1}^{\theta_2} d \ln(\theta) \langle \gamma_t \rangle, 
\end{eqnarray}
which measures the azimuthally-averaged, mean surface density interior to 
$\theta_1$ relative
to the mean in an annulus $\theta_1 < \theta < \theta_2$. The outer
radius of the control annulus was fixed at $\theta_2 = 4\farcm6$.
Displayed as well are the curves associated with the three 
isothermal-like surface mass distributions discussed above.

Based on the observed radial shear profile and corresponding 
$\zeta$ statistic,
it is clear that the azimuthally-averaged radial mass surface density at 
angular distances $\theta > 1^\prime$ is not steeper than isothermal 
($\theta^{-1}$) and in fact, is likely to be shallower, consistent with the 
flat light surface brightness profile.

\begin{figure}
\myplotone{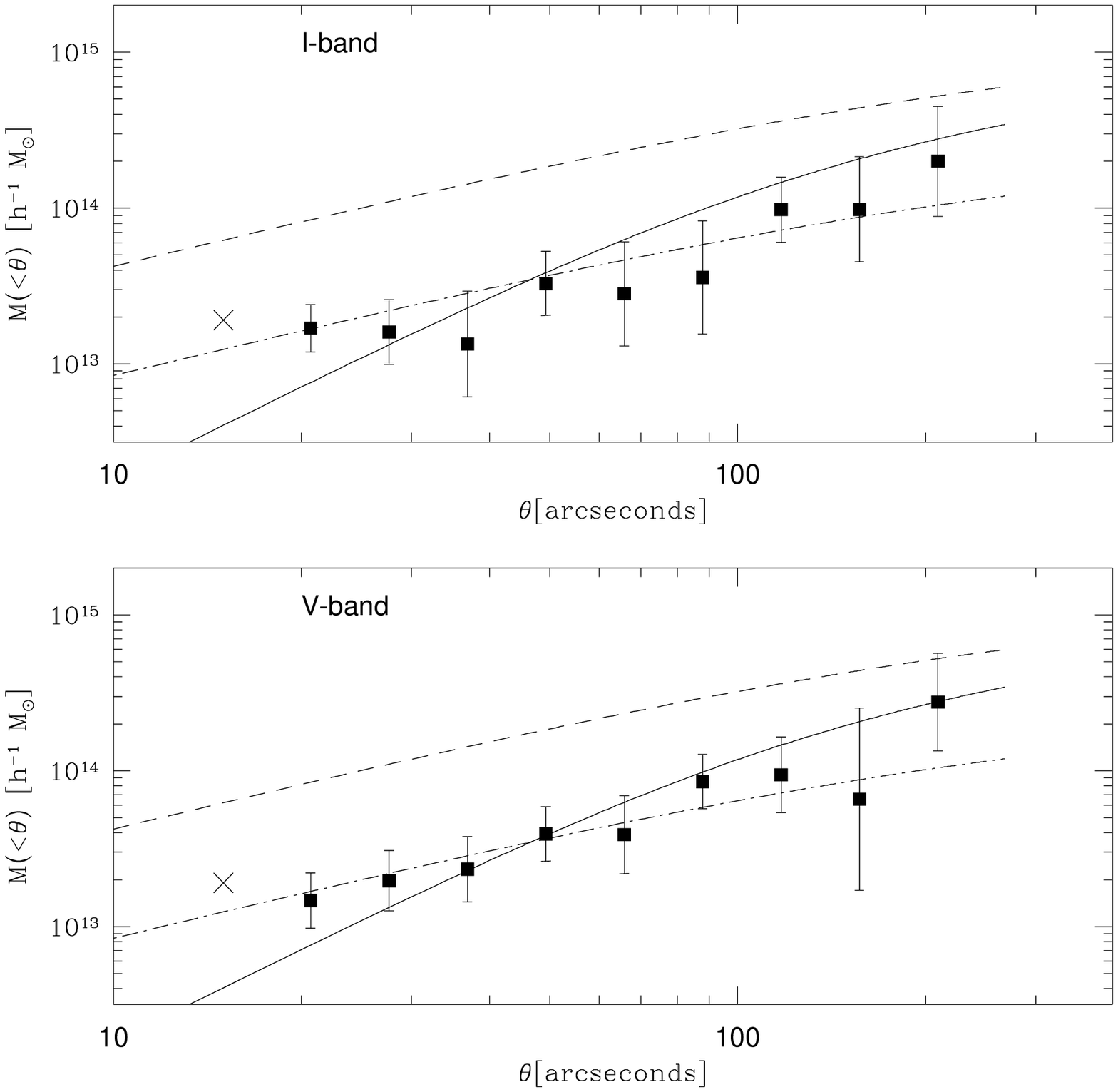}
\caption{The radial mass profiles for A2163 for the V- and I-band data (bottom
and top panels respectively) with the center chosen to be the giant
elliptical galaxy. The mass is a cumulative function, and hence 
the uncertainties on different points are correlated. The lines are the
model predictions with parameters as described in Figure 
\protect{\ref{fig:a2163_etprof}}. In these calculations, we have
corrected for
the mass in the control annulus to facilitate direct comparison with
the lensing masses derived using the $\zeta$ statistic. 
The ``x'' represents the mass
determined by assuming the arcs lie at the Einstein radius.
The errors on the weak lensing points were determined by adding the errors in
$\kappa$ and $\Sigma_{crit}$ in quadrature. The outer radius of the control
annulus was fixed at $4\farcm6$. The scale is 
$60^{\prime\prime} = 0.127 \Mpc$.}
\label{fig:a2163_MassvsR}
\end{figure}

To convert the $\zeta$ measurements into mass estimates, we 
estimate $\Sigma_{crit} = (4\pi G D_l \beta)^{-1} $, where
$D_l$ is the angular diameter distance to the lens and
$\beta = \hbox{max}( 0, \langle 1 - w_l /w_s 
\rangle )$.  In an Einstein de-Sitter universe with $\Omega = 1$, 
the comoving distance $w$ is defined as $w = 1 - 1/\sqrt{1+z}$.
We determine $\beta$ by an extrapolation of faint redshift
surveys (\cite{lilly93}; \cite{lilly95}; \cite{tresse93})
to fainter magnitudes and find,
for A2163 and our background galaxy selection criteria,  
$\Sigma_{crit} = (6.7 \pm 0.6) \times 10^{15}$~h~M$_\odot$/Mpc$^2$.
In Figure \ref{fig:a2163_MassvsR} we display the radial mass
profile for Abell~2163 with the giant elliptical galaxy taken as the 
cluster center. The results {}from the V-band data are displayed in the
bottom panel and the I-band results are shown in the top panel. 
We also mark the mass determined by MB
assuming the arcs  lie at the Einstein radius and plot the curves
corresponding to the three isothermal-like mass distributions.

Once again, the comparison of the derived mass profile and those corresponding
to the isothermal-like models must be undertaken with care.  As discussed
previously, the dilution of the shear signal in the central regions ($\theta
< 1^\prime$) results in the mass estimate in the central regions being biased
downward.  In addition, the mass estimates derived {}from the lensing analysis
are differential in nature --- the surface density at any
point is measured relative to the mean in the control annulus. Hence,
if the cluster mass distribution extends into the control region --- a 
likely scenario given that the field of view being analyzed is relatively 
small, the cluster light profile continues to rise and that 
the X-ray surface
brightness profile extends to $\theta\approx 15^\prime$
--- the lensing mass estimates will underestimate the true mass
by some significant amount.  We attempt to address this latter issue when we
make comparisons with the mass profile derived {}from the X-ray observations
(see \S \ref{sec:comparison}). Note, however, that the mass estimate of
MB is not subject to these uncertainties.

\begin{figure}
\myplotone{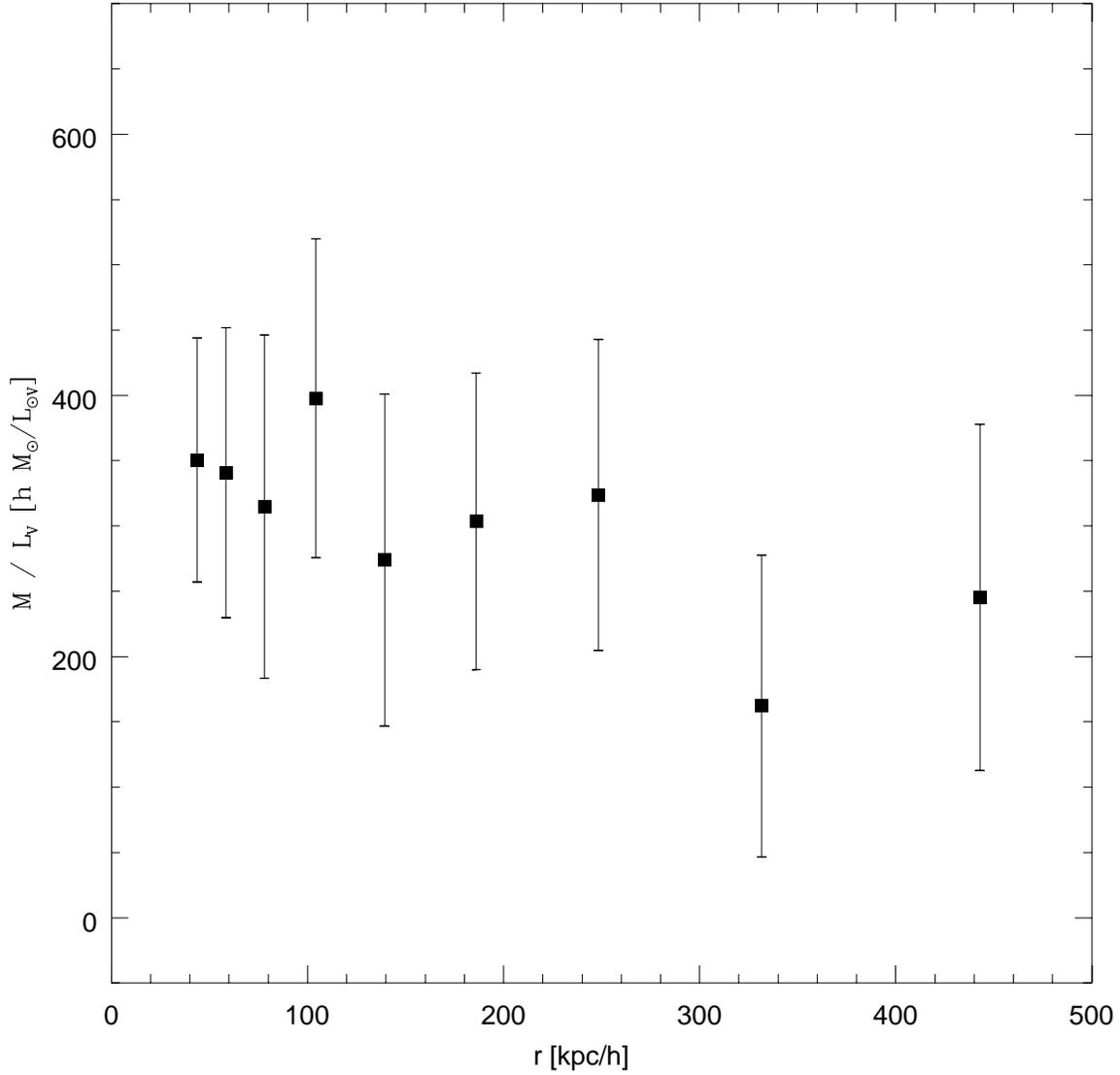}
\caption{The mass-to-light $M/L_V$ ratio for 
A2163 versus radius {}from the giant elliptical galaxy. The light estimate
comes {}from using all the light associated with galaxies 
detected in the image (including potential contributions
{}from galaxies not associated with the cluster). 
The mass-to-light ratio was formed by calculating the
mean surface mass and light densities as a function of radius, 
relative to the mean in the control annulus. Both quantities are cumulative
and thus the uncertainties on different points are correlated.}
\label{fig:a2163_MLvsR}
\end{figure}

Using the cluster light and the lensing mass profiles, we can compute the
cluster mass-to-light ratio as a function of radius. 
To account for the subtraction of the material in the control annulus
in the mass estimator, we form an analogous quantity, $\zeta_L$, for the 
light.  We calculate the mean light surface density at each radius relative
to the mean in the control region.  Under the assumption that mass
traces the light --- the cluster light and the lensing mass profiles are
consistent with this assumption, the ratio of the $\zeta$ values based on
the observed shear and $\zeta_L$ forms an unbiased estimate of the 
mass-to-light ratio in the cluster.  This ratio is plotted
in Figure \ref{fig:a2163_MLvsR}. The light estimate
comes {}from using all the light associated with the galaxies 
detected in the image (which, as we have 
discussed previously, probably includes a contribution
{}from galaxies not associated with the cluster). We find
a mass-to-light ratio of $M/L_V = (300 \pm 100) h M_\odot/L_{V\odot}$ at
$\theta\approx 2^\prime\approx 0.250 \Mpc$, with the radial profile
being consistent with a constant $M/L_V$ across the cluster.

\section{X-ray analysis}

\begin{figure}
\myplotfiddle{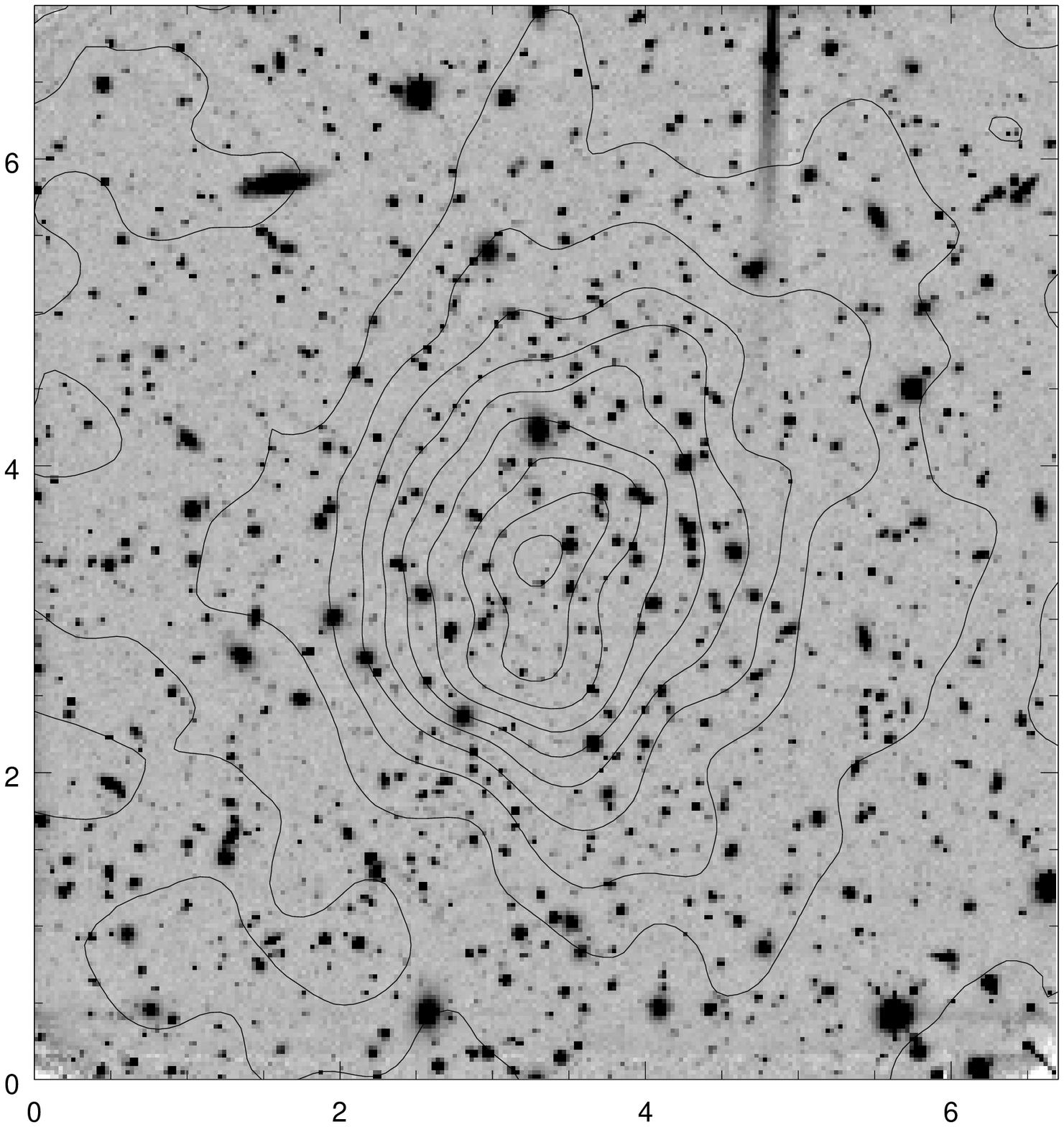} 
\caption{The ROSAT HRI
image of the X-ray emission (contours) placed on the
optical image of A2163. The contours are
smoothed with a Gaussian filter with scale
$12^{\prime\prime}$. The contour levels are
linearly spaced by $7\times10^{-8}$~counts/s/sq. arcsecond
with the maximum contour corresponding to
$8.9 \times 10^{-7}$~counts/s/sq. arcsecond.
North is to the right; East is up. The
angular scale is arcminutes ({\it or $127 \Mpc$}).}
\label{fig:xrayonimage} 
\end{figure}

Abell 2163 was observed by ROSAT for 12133s in
total with the PSPC (in two observations) and for
36194s with the HRI.
Figure \ref{fig:xrayonimage}
shows the HRI image of
A2163 superimposed on the rebinned V-band image of the cluster. The higher 
resolution HRI data 
confirm the results of the PSPC observation,
as discussed in EAB: The
centroid  of the X-ray emission is not associated
with any obvious optical counterpart -- the 
peak of the emission lies about $1^\prime$ west of the giant
elliptical galaxy with the arcs. 
The central region shows an elongation in
the east-west direction, roughly along the
direction defined by the two main concentrations in
the optical and total matter distributions. 
The X-ray morphology, however, appears
to be much more regular than the galaxy or total
mass morphology, displaying clear concentric isophotes, that are
elongated roughly 
in the same direction as the central part of the cluster.
As the HRI provides a much higher
spatial resolution it shows the  innermost part of
the cluster in more detail than the PSPC data.
Interestingly, there is an indication of bimodality in the inner arcminute.
This substructure, which was not resolved by the
PSPC, is aligned with the large scale elongation
axis. 

To quantify the total mass distribution we proceed
in the standard fashion. We fit an isothermal
$\beta$-model (\cite{jones84}), 
convolved with the instrument PSF, to the  surface
brightness distribution in the ROSAT/PSPC image.
This allows an
analytical  deprojection of the surface brightness
profile, $S(r)$, into a gas density distribution
$\rho(r)$: 
\begin{eqnarray}  
S(r) & = &
S_0\left(1+\frac{r^2}{r_c^2}\right)^{1/2-3\beta}+B \nonumber \\ 
\rho(r) & =
&\rho_0\left(1+\frac{r^2}{r_c^2}\right)^{-3\beta/2} \label{eqn:rhoofr}
\end{eqnarray} 
Formally, this inversion assumes that the gas is isothermal 
and spherically symmetric. However,
since the ROSAT sensitivity varies only very weakly with the emission
temperature (6\% {}from 2~keV to 10~keV) 
(\cite{jones92}; \cite{bohringer94}) 
the surface brightness fit is
still a very good representation of the gas density profile even though the
cluster is not isothermal.

Assuming a constant background, $B$, 
EAB obtained for the fit-parameters 
$\beta=0.620^{+0.015}_{-0.012}$ and $r_c = (153 \pm 10) \kpc \equiv 
1\farcm20\pm0\farcm075$. These parameter error estimates only take 
into account the statistical uncertainties on $B$.
We reexamine these results allowing for 
systematic errors in the background. Such 
errors can occur due to either improper vignetting
correction, the contribution of the support
structure of the X-ray telescope, or  due to the
contribution {}from unresolved point sources.
The latter maybe caused by the fact that in areas with large off-axis angles,
the PSF of X-ray telescopes
usually increases by a  substantial amount.  These
background variations of the PSPC data, derived by
EAB {}from a study of the fluctuation in the
counts, are of the order of 9\%. Including  these
uncertainties, by fitting the data with a
background $\pm$10\% and adding this systematic error in quadrature to the
error estimates obtained by EAB, we obtain new errors for $\beta$ and
for $r_c$, so that $\beta = 0.620 \pm 0.035$ and $r_c = (153 \pm 16) \kpc
\equiv 1\farcm20 \pm 0\farcm13$.

In addition to the PSPC data we analyzed the data of
the HRI observation. These provide a higher
resolution of the central area of the cluster,
which is relevant for comparison with the weak lensing analysis done here.
The HRI data also permit a test of how the
derived mass profile is affected by the limited
resolution of the PSPC (we expect the effect to be small since
the PSF, which is included in the above
analysis, has a FWHM of 20$^{\prime\prime}$ compared to the $1\farcm2$ 
core radius).
The downside is that the HRI detector has a higher
internal background, so that the cluster emission can
only be traced out to approximately $1 \Mpc$,
compared to more than $2 \Mpc$ with the PSPC.
Fitting the HRI data we obtain $\beta$ and $r_c$ values consistent with 
the PSPC
results: $\beta$= 0.57, and $r_c = 142 \kpc$ with  errors of 
$\simeq \pm $10\%. 
Due to the lower sensitivity and the higher
background of the HRI, these errors are larger than the errors for the PSPC 
fit including the background variations.

By combining the fits {}from the isothermal $\beta$-model of EAB
with the temperature profile,
and employing the assumptions of spherical symmetry and hydrostatic
equilibrium, we determine the total mass via the relation
\begin{equation}  M(<r)= -\frac{r^2kT}{G\mu
m_p}\left(\frac{d\rm{ln}\rho}{dr}
	+\frac{d\rm{ln}T}{dr}\right) .
\end{equation} 
In addition we estimate the gas
mass of the cluster by integrating the density given in equation
\ref{eqn:rhoofr}, under the assumption of spherical symmetry.
We discuss the uncertainties in the mass estimate 
introduced by the above assumptions in Section 3.1.  
Here, we simply note that at present, the uncertainties 
in the mass estimate are dominated by large uncertainties in
the temperature. A more detailed justification is given below.

Two different azimuthally averaged, radial temperature 
profiles of the ICM are considered for the 
determination of the total mass.
We first consider the results obtained by MMIYFT
{}from a combined fit  of the ASCA/GIS and SIS data:
11.0~keV to 14.1~keV 
 within the innermost $445 \kpc$ ,  8.6 to
14.2~keV for  $445 \kpc < r < 953 \kpc$,
and   2.9 to 4.9~keV for $953 \kpc < r < 2 \Mpc$ (see also 
Table \ref{tab:tofr}).

MMIYFT noticed that the overall GINGA best
fit temperature is higher than the one just mentioned. To probe the
effect of this, 
we considered a second temperature profile that comes {}from the 
combined ASCA/GINGA fit of MMIYFT.
The resulting increased temperature may resemble more the real temperature
profile of A2163 as in principle the GINGA~LAC
instrument is more reliable for  the
determination of such high temperatures, due to its
higher sensitivity especially at high energies. To
assess the possible impact of these systematic
differences (which are of the order of the statistical errors) 
on the mass estimate, we used the
following temperature profile: a temperature of
10.5-16.2~keV within  $445 \kpc$, to
8.6-18.4~keV between 445 and $953 \kpc$, and
2.9-5.5~keV up to a radius to $2 \Mpc$ (see also Table \ref{tab:tofr}). This
profile is centered on the best fit values deduced by
MMIFYT {}from their simultaneous fit of SIS, GIS and
GINGA data and the errors are conservatively set such
as to encompass all the allowed values found by
MMIFYT {}from the different instruments.  

\begin{table}
\begin{center}
\begin{tabular}{c|c|c}
r & T(r) & T(r)\\ \hline
$[\mbox{kpc}]$ & [keV] & [keV] \\
\hline 
0 - 445 & 11.0 - 14.1 & 10.5 - 16.2 \\
445 - 953  & 8.6 - 14.2 & 8.6 - 18.4 \\
953 - 2000 & 2.9 - 4.9 & 2.9 - 5.5 \\
\hline
\end{tabular}
\end{center}
\caption
{The temperature profile for A2163 used as input for our mass determination. 
The right panel
shows the temperature profile obtained by MMIYFT fitting ASCA and GINGA data
simultaneously. Our conservative errors (see also text) are centered on 
MMIYFT's best fit data. 
The left profile is 
obtained by MMIYFT using only ASCA data.
Using an updated version of the GIS effective area,
the results based only on ASCA data yield higher temperatures, more in 
agreement with
the combined fit (Markevitch, private communication).
}
\label{tab:tofr}
\end{table}

To account for these substantial uncertainties of the temperature
profile in the mass determination we use the method developed by Neumann \&
B\"ohringer (1995). It is based on a Monte-Carlo
technique that translates errors in temperature into
errors of the total mass.  This is critical as the
uncertainties in temperature are  the main source of
uncertainty in the final results, and the gradient
of  the temperature enters the mass equation. 
The method has the further advantage, over parametric
methods used in previous studies (EAB; MMIYFT) for this cluster, to be
free of any assumption on the radial potential shape. The
method calculates possible temperature
profiles within the given  temperature range
determined {}from the X-ray spectroscopy in
different rings and includes the additional constraint
that the resulting mass  profile has to increase with
radius. We use a stepwidth of  $160 \kpc$, and
calculate  $10^3$ physically allowed temperature
profiles.  Because the errors in temperature  are quite
high we limit the range in which the
temperature at a certain step in the profile, $T[i]$, is
randomly set.  To facilitate this, each temperature
$T[i]$ is restricted  to lie within a  window  with a
size of 2~keV, the center of the window having the
same relative distance to the upper and lower
boundary of the spectroscopically derived temperature
distribution as $T[i-1]$.
Even though the observed temperature profile shows a
strong gradient, the Monte-Carlo method 
produces
physically correct temperature profiles (e.g., increasing mass with increasing
radius) that cover
almost the whole range of the temperature profiles
allowed by the observations.  

Our derived mass profile for A2163 is in general agreement with those 
derived by EAB and MMIYFT (which relied on the mass estimation method
described by Hughes (1989), and Henry \et (1993)). 
There are some slight differences between our results with the previous
studies. For example, MMIYFT's mass estimates are a bit lower than our results
when renormalized to $H_0$=100, with values of
1-1.5$\times 10^{15} \; M_\odot$ at a radius of roughly $1 \Mpc$ 
(although we note that the estimates formally agree
within the experimental uncertainties). The difference probably 
arises due to the assumption of the shape of the total mass distribution 
combined with the observed steep temperature gradient.
We compare our mass estimates with those derived by EAB. At the upper
end of the region allowed by the statistical uncertainties, our mass
estimate is in agreement with EAB's. However, we permit a lower mass
value than EAB.
We ascribe this difference as being 
due to the different temperature profile EAB employed (their
temperature profile does not have such a steep temperature drop at the outer
boundary of the cluster, so that their lower limit on the mass is higher).

We also calculated the mass distribution using the $\beta$ fit 
values of the HRI data in combination with MMIYFT's temperature profiles.
In the outer regions the HRI results differ 
{}from the PSPC data. However, within  an inner radius of $1 \Mpc$
the integrated and along the line-of-sight
projected  total mass (necessary for the comparison with the weak lensing) 
of the cluster varies only by a
factor less than about 10\% for the best fit value in $r_c$ and $\beta$. 
This shows again that the different values for $\beta$
and $r_c$ do not seriously affect the mass profile. The total gas masses 
derived by the
HRI data agree within the error bars with the PSPC gas mass results
as could be expected {}from the agreement found for the surface
brightness profile parameters.

\begin{figure} 
\myplotone{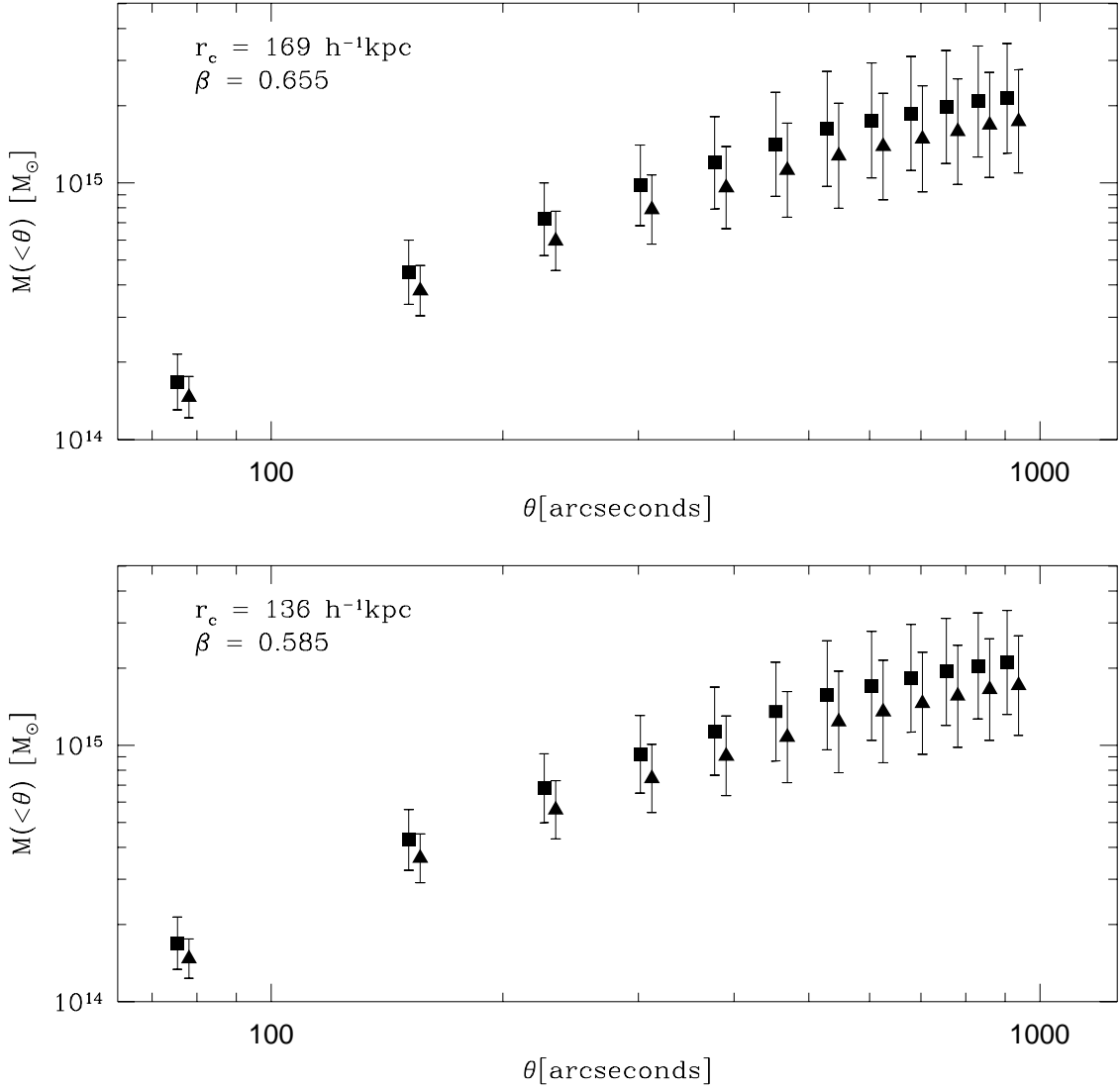}
\caption{The projected mass profiles determined {}from
the X-ray analysis. The temperature profiles are
{}from MMIYFT: the squares are with the temperature parameters
shown in column 2 of Table 1 while the triangles show the results
allowing for a higher central temperature (see column 3, Table 1).
The errors are 2$\sigma$
errors of the $10^3$ different randomly calculated
temperature distributions. We integrate the mass
profile out to a radius of $1920 \kpc$ over the
line-of-sight.} 
\label{fig:radialxraymass} 
\end{figure}

To facilitate a comparison between the weak lensing result and the X-ray 
results, we projected the radial mass profile 
derived {}from the Monte-Carlo technique  onto the
plane perpendicular to the plane of sight. 
The results are shown in Figure 
\ref{fig:radialxraymass}.
In the two panels we show the results obtained by choosing 
$\beta$, $r_c$ values
that lie at the extremes of the errors for the PSPC fit values.
The errors shown are $\pm2\sigma$, calculated for the $10^3$ simulations. The
effects of the uncertainties in the isothermal $\beta$-model are small,
changing the radial mass profile by 9\% at any radius, and 
the projected total mass by less than 4\%.
The effect of the variation in the
temperature profiles is larger, yielding a $\simeq
20$\% increase in the total mass estimates, when the
new temperature profile is used,  although the results
are consistent within the statistical uncertainties.

\begin{figure}
\myplotone{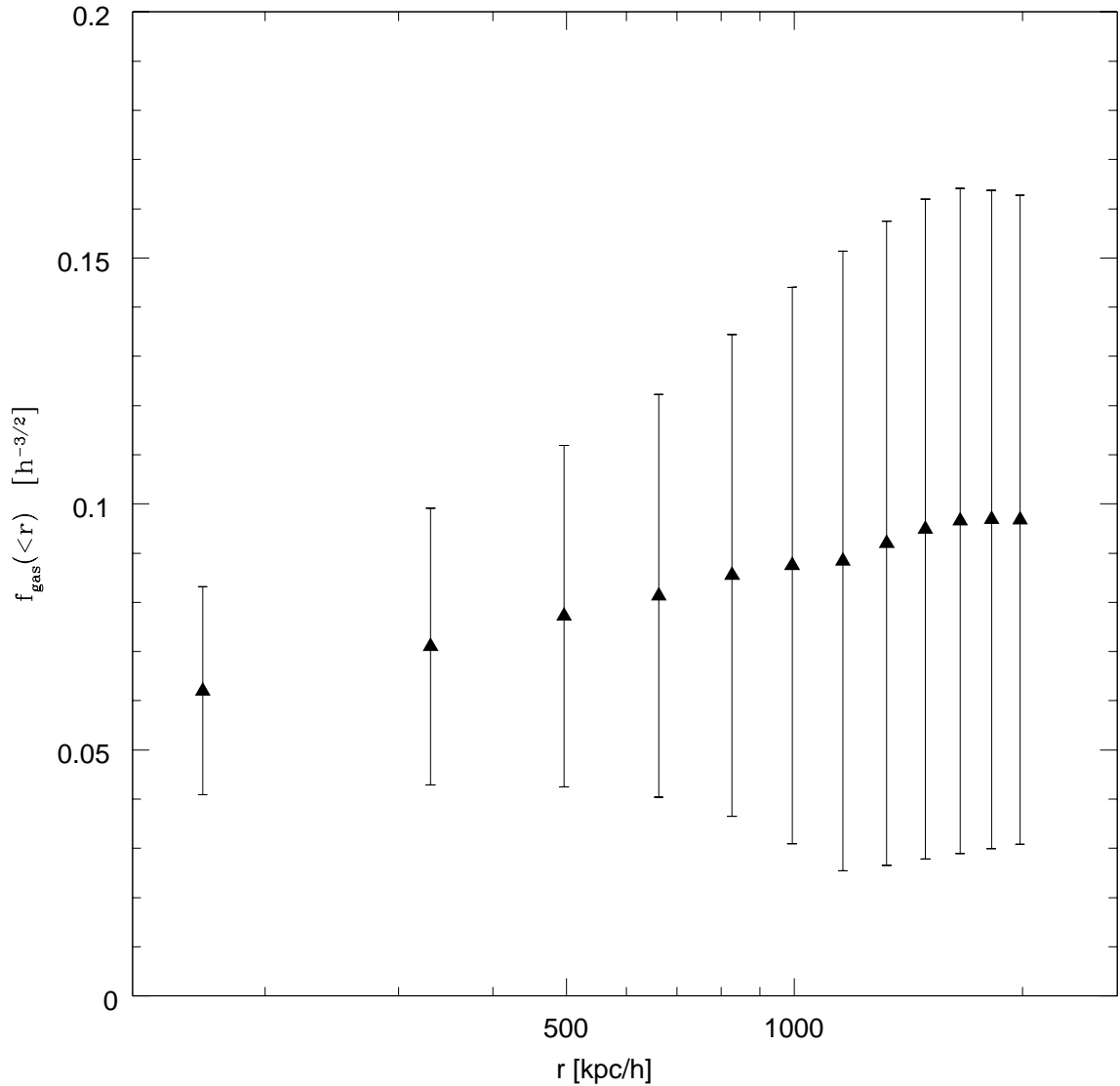}
\caption{The projected gas mass fraction as a function of radius {}from the 
X-ray emission centroid. The total mass estimate used was determined by
averaging over the mass estimates for the various $\beta$-fit parameters
and two temperature profiles. For the central gas density, we employed 
the best fit values from the different fits. The errors were determined
adding the uncertainties of the cumulative total and gas mass estimates
in quadrature and hence the uncertainties of different points are 
correlated.The projected gas mass fraction as a function of radius {}from the 
X-ray
emission centroid. The total mass estimate used is determined by
averaging over the mass estimates for the various $\beta$-fit parameters
and two temperature profiles. For the central gas density we take the best 
fit values for the different fits.}
\label{fig:gasfrac}
\end{figure}

Finally, in Figure \ref{fig:gasfrac} we plot the 
gas-to-total mass fraction for 
A2163 using only the X-ray data.
The gas mass determination based on ROSAT observations
is quite insensitive to the temperature.
To derive the gas mass profile, we employed the PSPC fit
with the errors in $r_c$ and $\beta$ quoted above, and the corresponding best 
fit values for the central density.  
Comparing the total mass
of the cluster with its gas mass yields a mean value of 
$M_{gas}/M_{tot} \simeq 0.07 \;h^{-3/2}$, with the statistical 
uncertainties allowing the range $0.02-0.16 \; h^{-3/2}$.

\subsection{Robustness of the X-ray mass estimates}

Given all the evidence that A2163 is not a well relaxed system, the 
question may arise as to how this will effect the X-ray mass determination
which depends on the
assumption of hydrostatic equilibrium. 
This is in fact a quite general
problem in the studies of clusters of galaxies:
Since the formation
time of clusters is comparable to the Hubble time, clusters are
essentially still forming today. In particular, in a high density
Universe, clusters are continuously accreting new material, quite often
in a discontinuous fashion (e.g., \cite{richstone92}; \cite{navarro95}).

Various numerical simulation studies have been undertaken to
determine whether the above assumptions introduce significant
uncertainties in the mass estimates.
In two recent simulation studies, Schindler (1996) and Evrard
\etal (1996) analyzed the validity of the mass
determination method used in X-ray analyses.
In these studies the mass was determined using the same procedure we employed
here, with the assumption that the gas is in hydrostatic
equilibrium in the gravitational potential of the cluster. 
When the derived mass profiles were compared with the 
true mass profiles, the results were found to be,
on the average, accurate within about 15 to 20\%, 
indicating that the mass determination {}from X-ray data 
is quite robust.
As pointed out by Schindler
(1996), larger deviations can occur particularly in those cases where a
major merger shock wave passes through the cluster -- the mass value
can be well off by a factor of two locally, near the position of the
shock wave -- and in cases of major substructural features that are not
properly accounted for.

There are some questions as to the accuracy of the simulations
(\cite{anninos96}).
The important point that we wish to make here is that, at present, 
the uncertainties 
introduced by deviations {}from hydrostatic equilibrium
and spherical symmetry are unlikely to be as large as 
the uncertainties in the temperature profiles. Indeed, we quantify this
large uncertainty by employing the Monte Carlo method to
determine the cluster mass. In this cluster, the uncertainty in the temperature
introduces a factor of $\simeq 2$ error in the mass estimate
both on large as well as small scales.
   
We consider below (see \S \ref{sec:speculation}) the possibility that the
cluster has undergone a recent merger and we are observing it in the
first  rebounce phase. In this case, the simulations of Schindler \& M\"uller
(1993) provide a numerical model to guide the interpretation of our
results. At the time of core passage of the two subcluster centers, the major 
thermalizing shock wave
is started {}from the center of the cluster. The time since core passage
can be estimated to be of the order of 0.5~Gyrs, 
based on the relative positions of the galaxy and mass concentrations.
In this time the major shock wave has traveled about 1~Mpc.
The X-ray mass profile is, however, derived for the whole cluster
volume and since this profile is found to be very smooth (and the
temperature profiles used here do not provide the resolution to resolve
the temperature structure of the shock wave  in any case, contrary to 
the resolution available in the simulations),
the mass results we determine here will be little affected by the local 
deviation at the shock front. 

\section{Comparison of the mass, light and gas distributions} 
\label{sec:comparison}        

The combined analysis of the optical, lensing and X-ray data 
permits a direct comparison of the distributions of the galaxies, 
the hot gas, and the gravitational mass in A2163. These 
distributions are shown in Figures \ref{fig:a2163_lightdistributions},
\ref{fig:a2163_massonimage} and \ref{fig:xrayonimage}, respectively.
The projected distribution of the galaxies, the light and the 
total mass are quite similar.  They all have extended, irregular
morphologies exhibiting a very strong east-west elongation
accentuated by two peaks. One of the maxima is located near the 
central giant elliptical galaxy about which arcs have been 
detected and the other is found between this galaxy and another
bright elliptical.  The similarity between the light and
the mass maps suggests that light is tracing the mass         
in this cluster.  This assertion can be further demonstrated by
a comparison of the azimuthally-averaged radial optical surface
brightness profile and the azimuthally-averaged radial surface
mass density profile.  Both have a similar radial dependence that is 
consistent with $l(\theta),\; \Sigma(\theta) \propto \theta^{-0.4}$, 
which is shallower than isothermal.   In Figure \ref{fig:a2163_MLvsR}, 
we show the mass-to-light ratio for the cluster and find that it is
constant with radius within the given uncertainties with an average
value of $M/L_V = (300 \pm 100) \;h M _{\odot} /L_{V \odot}$.

The lensing mass distribution was determined independently from the
V- and I-band data. Several interesting features arise from the shear
measurements and the inversion for the 2D mass profile. First, as
seen in Figure \ref{fig:a2163_etprof}, the radial shear profile is rather 
flat and is much weaker than that induced by a singular
isothermal profile characterized by a velocity dispersion comparable to that
observed for the cluster galaxies, even if one takes into consideration the
possibility that shear in the inner $1^\prime$ might be higher than observed.
On the other hand, a mass distribution 
also characterized by $\sigma=1680$~km/s but with a core radius comparable
to that observed in the X-ray produces less shear than observed, particularly
in the inner $1^\prime$ (and such a profile would not be able to 
produce the thin large arcs observed in the vicinity of the giant elliptical
galaxy.  Even if the cluster surface mass is flat over
the range $1^\prime < \theta < 2^\prime$, both the observed shear profile and
the thinness of arcs suggest that the cluster mass profile must steepen in
the central regions). A singular model that provides a reasonable fit to the 
measured shear corresponds to $\sigma=740$~km/s
although a singular isothermal model with $\sigma$ as high as $1000$~km/s
is also consistent with the observations, particularly if one takes into 
account the possible suppression of shear in the inner regions of the cluster
by contamination.  Indeed, under the simple assumption of a spherical,
isothermal mass distribution, a velocity dispersion of $\simeq 930$~km/s
would be required to form the Einstein radius at the observed arc
position.

The result of the flat, extended nature of the shear profile is that
the significance of the features seen in the 2D mass reconstruction is
not very high (especially in comparison with other clusters we have
studied). We do see some evidence for a bi-modal structure, with
each peak being resolved at the $\simeq 3\sigma$ level. The small scale
deviations between the 2D mass reconstructions seen in the
I- and V-band data can be partially attributed to a combination of seeing and
small number statistics. Moreover,
we wish to stress that if the cluster density profile consists of a
large, very extended, relatively flat structure, with small scale 
density substructures supposed on the flat plateau, then we would expect
to observe a small amplitude for the shear 
(in the limit of a constant surface density sheet, there is no shear 
generated). Hence, if this plateau 
extends right across our field of view, we would detect a weak shear signal 
and explain the relatively small significance of the mass peaks in 
the reconstructions. This is an observationally testable hypothesis with
larger angular scale optical observations.

In contrast to the projected galaxy, light and total matter 
distributions, the ROSAT/HRI X-ray surface brightness profile
has a much more regular appearance, and is a centrally peaked distribution.
The surface brightness distribution has one maximum located
between the two maxima of the optical light and projected
mass distributions.  The X-ray isophotes are elongated in
the east-west direction but the elongation is much less 
pronounced than that seen in the distribution of light and 
mass.  To some extent, a smoother distribution of the
X-ray emission is theoretically expected because if the 
gas is in hydrostatic equilibrium within the cluster's
gravitational potential, its distribution traces the 
potential, and the gravitational potential always has a 
smoother structure than the mass distribution 
(e.g., \cite{binney87}; \cite{buote96}).
             
Like the surface mass density and the optical
surface brightness profiles, the radial X-ray surface 
brightness profile is very shallow in the center but falls off
more steeply at larger radii.  
The X-ray surface brightness 
profile is weighted by an integration of the gas
density squared in the line of sight. Accounting for this 
square dependence, one finds a radial dependence
of the projected gas density at large radii of $\propto 
\theta^{-0.8}$.  In Figure \ref{fig:gasfrac}, we 
show the gas mass fraction as a  function of radius
and find a mean value of $M_{gas}/M_{tot} \simeq 
(0.07 \pm 0.03)\; h^{-3/2}$.

\begin{figure}
\myplotone{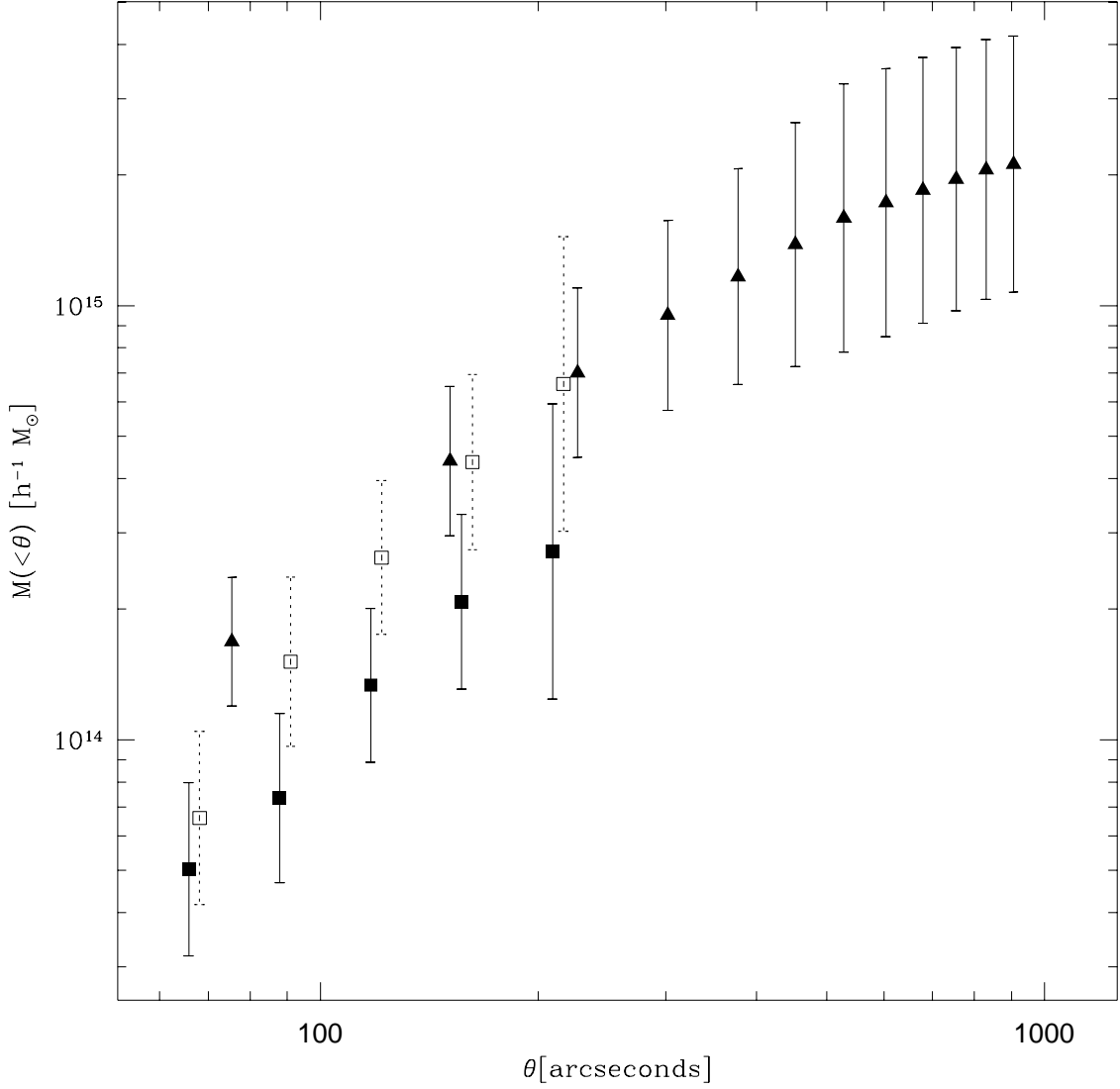}
\caption{The radial mass profiles determined {}from the X-ray and lensing
analysis. The triangles display the total mass profile determined {}from 
the X-ray data. The solid squares are the weak lensing estimates. The open
squares are the lensing estimates ``corrected'' for the mean surface
density in the control annulus determined {}from the X-ray data. The 
conversion {}from 
angular to physical units is $60^{\prime\prime} = 0.127 \Mpc$.
}
\label{fig:xraymasscomp}
\end{figure}

In order to compare the X-ray and lensing mass results we recalculated
the lensing mass profiles by taking the X-ray centroid as the center.
We also averaged the two different X-ray mass profiles (shown in Figure
\ref{fig:radialxraymass}) 
corresponding to the two extreme data sets for $\beta$, $r_c$, and
the two temperature profiles. The lensing and X-ray mass profiles are
compared in Figure \ref{fig:xraymasscomp}. The data sets only overlap out 
to a radius of $200'' \simeq 500 \;\kpc$  to which the lensing analysis is
limited. In the overlap region the lensing mass estimates are
systematically lower by a factor of $\simeq 2$ than the X-ray
results, although formally the results are consistent with each other
given the substantial uncertainties.

As we discussed in \S \ref{sec:lensing}, one important consideration 
to bear in mind is that the lensing mass estimate may be biased 
downward if the cluster mass extends into the control annulus used to 
calculate the $\zeta$-statistic.  For A2163, this is a likely 
scenario as our optical observations cover a relatively small field 
of view.  The shallow surface mass density profiles and the optical 
surface brightness profiles both  suggest that the cluster is quite 
extended.  In addition, the cluster X-ray emission has been traced 
to $2.2 \;\Mpc$.  The only accurate, model-independent way to avoid
the bias is to observe 
and analyze the weak lensing distortions in the cluster out to 
larger radii. In the absence of necessary observations, 
we can attempt to estimate the magnitude of the bias 
affecting the lensing data based on the analytic spherically 
symmetric models fit to the observed radial shear profile 
(see Figure \ref{fig:a2163_etprof}).  We find that the actual 
projected cluster mass could be anywhere {}from a factor $1$--$2$ 
higher, depending on the assumed model for the mass distribution.  
Alternatively, we can start with the null hypothesis 
that the lensing and X-ray mass estimates should agree, determine, 
based on the X-ray mass profile, the mean surface mass density in 
the control annulus and correct the lensing $\zeta$ estimate at 
every radius by this amount.  The results of applying such a 
``correction'' is displayed in Figure \ref{fig:xraymasscomp}.  As 
expected, the weak lensing mass estimates increase and the agreement
between the lensing and X-ray results improves.  
         
Having found consistency between the two data sets, we can         
consider more closely the cluster mass-to-light ratio and the         
cluster gas mass  fraction.  If on large scales, the         
dark matter and the baryons are not segregated, and the clusters          
of galaxies accrete gas (baryons) and dark matter in an          
indiscriminant way during their formation, and  the         
efficiency of galaxy formation is the same in all          
environments, then clusters of galaxies can be thought of as         
representative elements of the Universe.  The cluster          
mass-to-light ratio and the gas mass fraction, therefore,          
should equal their respective universal values.  Comparing         
the mass-to-light ratio of $M/L_V = (300 \pm 100) \;h          
M_\odot/L_{V\odot}$ for A2163 with mean value for the local         
Universe (see \cite{carlberg95}) implies a value of         
$\Omega \simeq 0.3 \pm 0.1$.  Similarly, comparing A2163's         
gas mass fraction of $M_{gas}/M_{tot} \simeq (0.07\pm 0.03) \;         
h^{-3/2}$ with the universal mean baryon fraction of $f_b \leq         
0.015 \;(\Omega h^{2})^{-1}$ predicted by nucleosynthesis         
(\cite{walker91}) also implies $\Omega \leq 0.4 \;h^{-1/2}$.         
Both estimates point  towards a low density Universe provided          
assumptions noted above are valid.

\section{Interpretation of the dynamical state of the cluster}
\label{sec:speculation}
 
Taken jointly, the differences in the distributions of the cluster 
light, gas and dark matter strongly suggest that far {}from being 
a relaxed cluster, A2163 is very much a system in the process 
of formation. This conclusion was noted previously by EAB {}from a study of 
the overall
galaxy distribution and the X-ray morphology of the PSPC image. For example,
one striking feature in A2163 is that the X-ray emission centroid
is so significantly devoid of any optical counterpart.
The ``cluster in formation'' hypothesis is further supported by the cluster's
very extended velocity histogram,  a high velocity dispersion and the fact
that coordinate-redshift galaxy distribution shows significant substructure 
(Soucail \etal 1996).  The extended velocity histogram and the high velocity
dispersion is most likely due to velocity distributions of infalling 
subclusters overlapping in velocity space.  The value of 
$\beta_T=\sigma^2/(kT/\mu m_p)=1.3$--$1.5$ is larger than
the canonical value of 1 and the values found in the range 1 to 1.2
for fairly quiescent clusters indicating that A2163 is not in a
relaxed dynamical state.
Furthermore, the presence of a luminous extended
radio halo (\cite{herbig95}) can be taken as yet another signature of 
the cluster having suffered
a recent merger event.
Very few clusters have large radio halos associated with them and 
all those that  do (e.g., A2256, A2255 or Coma) show signatures of recent 
merger (\cite{tribble93}; \cite{burns94}; \cite{briel94}; 
\cite{feretti96}).
Theoretical arguments further strongly favor the extended radio halo/merger 
connection (see \cite{bohringer92}; \cite{bohringer95}).
Finally, evidence of 
complex, small-scale 
temperature variations in the central regions of the cluster 
described by Markevitch \etal (1994) as well as the extremely steep drop in the
the mean gas temperature between the central regions and the outer regions
(MMIYFT), provide 
further evidence that the cluster, as a whole, is unrelaxed.  
Other clusters with temperature substructure such as
A754 (\cite{henry95}) and A2256 (\cite{briel91}; \cite{briel94}; 
\cite{roettiger95}) show evidence of recent/ongoing merger.

In the picture described above, the two major peaks in the light 
distribution probably
correspond to the two most dominant subclusters that are in the process of
merging (see also EAB). 
One scenario that describes the relative dark matter and gas
distributions is that the two colliding subunits of the cluster have
already passed through each other and are on their second
turn-around trajectory. In this interpretation the two maxima of the
galaxy distribution may be relics of the central regions of the two
former subclusters. The hypothesis is substantiated by comparison with
numerical simulations of cluster formation. 
In the N-body/hydrodynamical simulations by Schindler \& M\"uller
(1993) for example the central regions of the two subclusters are
traced by distinct maxima in the mass particles and the X-ray distribution
during most of the first infall period. During the rebounce phase,
after the central passage, the former subcluster centers are still
recognizable as clumps of mass particles for some time while the X-ray
surface brightness distribution appears smooth and not multi-peaked. At
this state the simulated clusters resemble very much A2163 in
appearance. Another well studied cluster which is most likely found in
a similar state is A2255. In this case the X-ray peak is
separated {}from the two dominant central galaxies and the cluster shows
an elongation in the central region and an overall elliptical shape
(\cite{burns94}; \cite{feretti96}). {}From simulation of the
cluster merger carried out to match and explain the morphology of A2255,
Burns \etal (1994) concluded that the cluster is in a rebounce phase.

This comparison of the observational features of A2163 with results from
simulations can of course not provide a unique identification of the
dynamical state of the cluster. In this case they seem to provide a 
very plausible scenario, however, which gains further support for example 
by the presence of the radio halo. It will be quite difficult, on the other 
hand, to substantiate this picture purely by more observational data.

\section{Summary} \label {sec:summary}

We have studied optical, gas and dark matter distributions in the cluster 
of galaxies Abell~2163.
We have traced the cluster galaxy light 
distribution over a $7' \simeq 1 \Mpc$ field centered on the dominant
central galaxy. Using the weak gravitational distortion of 
background galaxies and correcting for systematic effects that can bias
the galaxy shapes, we have mapped the dark matter distribution in the 
cluster over the same scale. Combining the observed X-ray surface brightness
profile {}from ROSAT/PSPC observations with the spectroscopically
determined temperature profiles, we have estimated the projected total
mass and gas mass distribution to a radius of $\simeq 2 \Mpc$. 
In the present case, we find agreement at the $2\sigma$ level between the raw
weak lensing and X-ray inferred masses,  with the X-ray mass determinations 
being a factor  of $\simeq 2$ higher. Correcting for matter in the control
annulus, as determined {}from the X-ray observations, yields a much
better agreement between the two results.
The extended nature of the mass distribution is consistent
relatively weak shear profile that was measured in this cluster. 
The extended and clumpy nature of the cluster galaxy light distribution, 
the broad cluster galaxy velocity histogram, the high value of
$\beta_T$, the extended luminous radio halo, and the 
irregular mass distribution inferred {}from the weak lensing analysis, 
suggests that A2163 is a cluster in the process of formation.

The observations presented here have been a useful check of the consistency
between the two mass determination methods and have enabled us to
speculate on the dynamical state of the cluster.
The somewhat unsatisfactory aspect of this is that, at least for
the lensing analysis, the mass determinations were confined to a relatively
small physical radius {}from the cluster center. Clearly, any conclusions
about the universality of the gas mass fraction and mass-to-light ratio are
subject to interpretations regarding the physics of clusters centers, and
how representative a remarkable cluster like A2163 is of the Universe
as a whole. A more pleasing comparison should, and now could, be done at 
over larger scales. Indeed, the technology now exists for
such large field lensing observations with the MOCAM 14$^\prime$ 
and the UH $\simeq 0.5^\circ$ cameras at CFHT. 
One exciting example is that, in hierarchical clustering scenarios, clusters 
of galaxies tend to form at intersections of filaments.  In clusters such as
A2163, which appear to be very much in formation, it may be possible to use
large field observations of the weak lensing to detect filaments along which 
mass is flowing into the cluster. With the types of
observations possible with the current generation of instruments, a 
wide range of outstanding issues can be probed regarding the dark matter
content and distribution in the Universe.

\acknowledgments{We would like to thank Sabine Schindler and Rien van de 
Weygaert for many enlightening discussions. It is a pleasure to acknowledge 
the valuable assistance of Richard Griffiths and the MDS team. 
We also gratefully acknowledge the redshift data and extrapolation to faint 
magnitudes provided by Simon Lilly, Caryl Gronwall, and David Koo. AB 
gratefully acknowledges support from The Dudley Observatory.}


\clearpage

\begin{thebibliography}{}

\bibitem[Abell \etal 1989] {abell89} Abell, G.O., Corwin, H.G., \& 
	Olowin, R.P. 1989, \apjs, 70, 1

\bibitem[Anninos \& Norman 1996]{anninos96} Anninos, P., \&
	 Norman, M. L. 1996, \apj, 459, 12

\bibitem[Arnaud \etal 1994] {arnaud94} Arnaud, M , Elbaz, D., B\"ohringer, H.,
	Soucail, G., \& Mathez, G. 1994, in New Horizons of X-ray Astronomy,
	eds. F. Makino \& T. Ohashi (Tokyo: Universal Academy Press), 537

\bibitem[Arnaud \etal 1992] {arnaud92} Arnaud, M., Hughes, J.P., Forman, W.,
	Jones, C., Lachieze-Rey, M., Yamashita, K., \&
	Hatsukade, I. 1992,  \apj, 390, 345

\bibitem[Babul \& Katz 1993]{babul93} Babul, A., \& Katz, N. 
	1993, \apjl, 406, 51

\bibitem[Bessell \& Brett 1988]{bessell88} Bessell, M.S. \& Brett, J.M.
	1988, PASP, 100, 1134

\bibitem[Binney \& Tremaine 1987]{binney87} Binney, J., \& Tremaine, S. 
	1987, Galactic Dynamics, (Princeton: Princeton University Press)

\bibitem[B\"ohringer 1995]{bohringer95} B\"ohringer, H. 1995,  in Reviews 
	in Modern Astronomy 8: Cosmic Magnetic Fields,
	ed.  G. Klare (Springer Verlag), 259

\bibitem[B\"ohringer 1994]{bohringer94} B\"ohringer, H. 1994, 
	in Proceedings of the NATO Advanced Study Institute on 
	Cosmological Aspects of X-ray clusters of Galaxies,
	ed. W. Seitter (Boston: Kluwer Academic Publishers), 123

\bibitem[B\"ohringer \etal 1992]{bohringer92} B\"ohringer, H., 
	Schwarz, R.A., Briel, U.G., Voges, W., Ebeling, H.,
	Hartner, G., \& Cruddace, G., 1992, in Clusters and Superclusters of
	Galaxies, ed. A.C. Fabian, (Dordrecht: Kluwer Academic Publishers),
	 71

\bibitem[Blumenthal \etal 1984]{blumenthal84} Blumenthal, G.R., Faber, S.M., 
	Primack, J.R., \& Rees, M. 1984, Nature, 311, 517

\bibitem[Briel \& Henry 1994]{briel94} Briel, U.G., \& Henry, J.P. 1994,  
	Nature, 372, 439

\bibitem[Briel \etal 1991]{briel91} Briel, U. G., 
	Henry, J.P., Schwarz, R., B\"ohringer, H., 
	\& Ebeling, H. 1991, \aap, 246, L10-13

\bibitem[Buote \& Canizares 1996]{buote96} Buote, D., \& Canizares, C.
	1996, \apj, 457, 565

\bibitem[Burns \etal 1994]{burns94} Burns, J.O., Roettiger, K., Ledlow, M. \&
	Klypin, A. 1994, \apj, 427, L87

\bibitem[Carlberg \etal 1995]{carlberg95} Carlberg, R., Yee, H.K.C., 
	Ellingson, E., Abraham, R., Gravel, P., Morris, S., \& 
	Pritchet, C.  1995, preprint

\bibitem[Coleman \etal 1980] {coleman80} Coleman, G.D., Wu, C.C., \& 
	Weedman, D.W. 1980, \apj, 43, 393

\bibitem[David 1995]{david95} David, L.P., Jones, C., \& Forman, W. 
	1995, \apj, 445, 578

\bibitem[Davis 1990]{davis90} Davis, L. 1990, private communication

\bibitem[Efstathiou 1995]{efstathiou95} Efstathiou, G. 1995, in 
	Les Houches Lectures on Galaxy Formation,
	Elsevier Science Publications, Netherlands 

\bibitem[Elbaz, Arnaud \& B\"ohringer 1995] {elbaz95} Elbaz, D., Arnaud, M.,
	\& B\"ohringer, H. (EAB) 1995, \aap, 293, 337

\bibitem[Evrard, Metzler \& Navarro 1995]{evrard95} Evrard, A., 
	Metzler, C., \& Navarro, J. 1995, preprint

\bibitem[Fahlman \etal 1994]{fahlman94} Fahlman, G., Kaiser, N., Squires, G.,
	\& Woods, D. 1994, \apj, 437, 56 

\bibitem[Feretti \& Giovannini 1996]{feretti96} Feretti, L., \& 
	Giovannini, G. 1996, Proc of the IAU Symposium 175
	on extragalactic radio sources, (in press)

\bibitem[Fort \& Mellier 1994]{fort94} Fort, B. \& Mellier, Y. 1994, 
	A\&AR, 5, 239

\bibitem[Henry \& Briel 1995]{henry95} Henry, J.P. \& Briel, U.G. 1995, \apj,
	443, L9

\bibitem[Henry, Briel \& Nulsen 1993]{henry93} Henry, J.P., Briel, U.G., \& 
	Nulsen, P.E.J. 1993, \aap, 271, 413

\bibitem[Herbig \& Birkinshaw 1995] {herbig95} Herbig, T., \& Birkinshaw, M.
	 1995, BAAS, 26, 1403

\bibitem[Holzapfel \etal 1996]{holzapfel96} Holzapfel,W.,  Arnaud, M., 
Ade, P., Church, S., Fischer,
M., Mauskopf, P., Rephaeli, Y., Wilbanks, T.M., Lange, A. 1996, \apj, 
submitted

\bibitem[Hughes 1989]{hughes89} Hughes, J.P. 1989, \apj, 337, 21

\bibitem[Jones \& Forman 1984]{jones84} Jones, C., \& Forman, W. 1984, 
	\apj, 276, 38

\bibitem[Jones \& Forman 1992]{jones92} Jones, C., \& Forman, W. 1992,
	{\it Clusters \& Superclusters of Galaxies}, ed. A.C.~Fabian,
	(Dordrecht: Kluwer Academic Publishers), 49
 
\bibitem[Kaiser \& Squires 1993]{ks93} Kaiser, N., \& Squires, G. 1993,
 \apj, 404, 441

\bibitem[Kaiser, Squires \& Broadhurst 1995] {kaiser95} Kaiser, N., 
	Squires, G., \& Broadhurst, T. 1995, \apj, 449, 460

\bibitem[Kneib \etal 1995]{kneib95} Kneib, J.P., Mellier, Y., Pello, R., 
	Miralda-Escud\'e, J., Le Borgne, J.-F.,
	B\"ohringer, H., Picat, J.-P. 1995, A\&A, 303, 27

\bibitem[Landolt 1992]{landolt92} Landolt, A. U.  1992, \aap, 104, 340

\bibitem[Lilly 1993]{lilly93} Lilly, S.J. 1993, \apj, 411, 501

\bibitem[Lilly 1995]{lilly95} Lilly, S.J. 1995, private communication.

\bibitem[Loewenstein 1996]{loewenstein96} Loewenstein, M. 1996, in proceedings
        of the 2nd international conference sponsored by UCLA on {\em Sources
        and Detection of Dark Matter in the Universe}, in press

\bibitem[Markevitch \etal 1994]{MYFT94}  Markevitch, M, Yamashita, K., 
	Furuzawa, A., \& Tawara, Y.  1994, \apj, 436, L71.

\bibitem[Markevitch \etal 1995]{MMIYFT95}  Markevitch, M, Mushotsky, R., 
	Inoue, H., Yamashita, K., Furuzawa, A., \& Tawara, Y.  (MMIYFT) 
	1995, \apj, accepted

\bibitem[Miralda-Escud\'e \& Babul 1995]{escude95} Miralda-Escud\'e, J. \& 
	Babul, A. 1995, \apj, 449, 18

\bibitem[Navarro \etal 1995]{navarro95} Navarro, J., Frenk, C.S., 
	\& White, S.D.M. 1995, \mnras, 275, 720

\bibitem[Neumann \& B\"ohringer 1995]{neumann95}Neumann, D.M., \& 
	B\"ohringer, H. 1995, \aap, 301, 86

\bibitem[Odewahn \etal 1992]{odewahn92} Odewahn, S.C., Bryja, C., \& 
	Humphreys, R.M. 1992, PASP, 104, 553

\bibitem[Richstone \etal 1992]{richstone92} Richstone, D., 
	Loeb, A., \& Turner, E.L. 1992, \apj, 393, 477

\bibitem[Roettiger, Burns \& Pinkney 1995]{roettiger95} Roettiger, K., 
	Burns, J.O. \& Pinkney, J. 1995, \apj, 453, 634

\bibitem[Sarazin 1986]{sarazin86} Sarazin, C. 1986, Rev. Mod. Phys., 58, 1

\bibitem[Schindler 1996]{schindler96} Schindler, S. 1995, \aap, 305, 756

\bibitem[Schindler \& M\"uller 1993]{schindler93} Schindler, S., \&
	M\"uller, E.  1993, \aap, 272, 137

\bibitem[Smail \etal 1995]{smail95} Smail, I., Ellis, R.S., Fitchett, M.J., 
	\& Edge, A.C. 1995, \mnras, 273, 277

\bibitem[Soucail \etal 1996] {soucail96} Soucail, G., Arnaud, M., 
	\& Mathez, G. 1996, in preparation

\bibitem[Squires \& Kaiser 1995]{sk95} Squires, G., \& Kaiser, N. 1995, 
	submitted to ApJ

\bibitem[Squires \etal 1996] {squires96} Squires, G., Kaiser, N., Babul, A.,
	Fahlman, G., Woods, D., Neumann, D. M., \& B\"ohringer, H. 1996,
	\apj, accepted

\bibitem[Stetson \& Harris 1988]{stetson88} Stetson, P. B. \& 
	Harris, W.E. 1988, \aap, 96, 909

\bibitem[The \& White 1987]{the97} The, L.S., \& White, S.D.M. 
	1986, AJ, 92, 1248

\bibitem[Tresse \etal 1993]{tresse93} Tresse, L., Hammer, F., le Fevre, O., 
	\& Proust, D. 1993, \aap, 277, 53

\bibitem[Tribble 1993]{tribble93} Tribble, P. 1993, \mnras, 263, 31

\bibitem[Tyson \etal 1990]{tyson90} Tyson, J., Valdes, F. \& 
	Wenk, R. 1990, \apjl, 349, L19

\bibitem[Walker \etal 1991]{walker91} Walker, T.P., Steigman, G., Schramm, 
D.N., Olive, K.A., \& Kang, H. 1991, \apj, 378, 186

\bibitem[White \& Fabian 1995]{white95} White, D. A., \& 
	Fabian, A. C. 1995, \mnras, 273, 72

\bibitem[White 1992]{white92} White, S. D. M. 1992, in
	Clusters and Superclusters of Galaxies, ed. A.C. Fabian, 
	(Kluwer: Dordrecht), 17

\bibitem[White \etal 1993]{white93} White, S.D.M., Navarro, J.F., 
	Evrard, A.E., \& Frenk, C. S. 1993, Nature, 366, 429

\bibitem[Wilbanks \etal 1994] {wilbanks94} Wilbanks, T.M., Ade, P.A.R., 
	Fischer, M.L., Holzapfel, W.L., \& Lange, A.E. 1994, \apj, 427, L72

\end{thebibliography}
\end{document}